\documentclass[preprint]{jfm}

\usepackage{graphicx}
\usepackage{newtxtext}
\usepackage{newtxmath}
\usepackage{natbib}
\usepackage{hyperref}
\usepackage{dcolumn}% Align table columns on decimal point
\usepackage{bm}% bold math
\usepackage{subcaption}
\usepackage{setspace}
\usepackage{adjustbox}
\usepackage{array}
\usepackage{caption}
\usepackage{booktabs}
\usepackage{array}
\usepackage{multirow}
\usepackage{amsmath} % for advanced math typesetting

\usepackage{esvect}  % for fancy vector arrows
\usepackage{soul}
\usepackage{float}
\hypersetup{
    colorlinks = true,
    urlcolor   = blue,
    citecolor  = black,
}

\newcommand{\RomanNumeralCaps}[1]
\title{Hydrodynamically Beneficial School Configurations in Carangiform Swimmers: Insights from a Flow-Physics Informed Model
}% 

\author{Ji Zhou\aff{1},
Jung-Hee Seo\aff{1}%
\and Rajat Mittal\aff{1}%
\corresp{\email{mittal@jhu.edu}}
}
\affiliation{%
 \aff{1}Department of Mechanical Engineering, Johns Hopkins University, Baltimore, MD 21218, USA
}%
\begin{document}
\maketitle

\begin{abstract}
Researchers have long debated which spatial arrangements and swimming synchronizations are beneficial for the hydrodynamic performance of fish in schools. In our previous work (Seo and Mittal, \emph{Bioinsp. Biomim.}, Vol. 17, 066020, 2022), we demonstrated using direct numerical simulations that hydrodynamic interactions with the wake of a leading body-caudal fin carangiform swimmer could significantly enhance the swimming performance of a trailing swimmer by augmenting the leading-edge vortex (LEV) on its caudal fin. In this study, we develop a model based on the phenomenology of LEV enhancement, which utilizes wake velocity data from direct numerical simulations of a leading fish to predict the trailing swimmer's hydrodynamic performance without additional simulations. This approach enables a comprehensive analysis of the effects of relative positioning, phase difference, flapping amplitude, Reynolds number, and the number of swimmers in the school on thrust enhancement. The results offer several insights regarding the effect of these parameters that have implications for fish schools as well as for bio-inspired underwater vehicle applications.
\end{abstract}
% \begin{description}
\begin{keywords}
Authors should not enter keywords on the manuscript, as these must be chosen by the author during the online submission process and will then be added during the typesetting process (see \href{https://www.cambridge.org/core/journals/journal-of-fluid-mechanics/information/list-of-keywords}{Keyword PDF} for the full list).  Other classifications will be added at the same time.
% Leading-Edge Vortex, Fish Schools, Swimming Wakes, Carangiform Swimmers, Immersed Boundary Method, Computational Fluid Dynamics.
\end{keywords}
%\tableofcontents
\section{\label{sec:background}Background}
The collective behavior of schooling fish is not only visually striking, it also presents a complex interplay of hydrodynamics, ethology, and environmental adaptation. Schooling and collective swimming provide several advantages, including improved foraging efficiency \citep{pitcher1982fish,pavlov2000patterns}, predator avoidance \citep{shaw1962schooling,larsson2012why,pavlov2000patterns}, stealth \citep{zhou2024effect}, and, notably, hydrodynamic efficiency \citep{weihs1973hydromechanics,partridge1979evidence,pavlov2000patterns,zhou2023effect}. The improvement in swimming efficiency has garnered significant attention for its implications for energy conservation during locomotion \citep{weihs1973hydromechanics,pavlov2000patterns,liao2007review,timm2024multi}. Specifically, fish within a school can harness the vortices generated by other fish, thereby reducing their energy expenditure \citep{seo2022improved,verma2018efficient,li2020vortex,taguchi2011rainbow,guo2023vortex}. This phenomenon has driven extensive research into the spatial configurations, tailbeat synchronization, and flow dynamics that govern schooling behavior \citep{seo2022improved,partridge1980three,maertens2017optimal,verma2018efficient,pan2024combining}.

The study of fish schooling has advanced significantly in recent decades, moving from foundational qualitative observations \citep{weihs1973hydromechanics,partridge1979evidence,partridge1980three} to sophisticated experimental \citep{shaw1962schooling,abrahams1985risk,becker2015hydrodynamic,marras2015fish,ashraf2016synchronization,newbolt2019flow,mckee2020sensory,wei2022passive,zhang2024collective} and computational \citep{maertens2017optimal,verma2018efficient,hang2022active,zhou2022complex,seo2022improved,zhou2024effect,pan2024unraveling,zhou2025hydrodynamic} investigations.  Early studies focused primarily on observable patterns and school formations, offering limited insights into the intricate fluid mechanics at play. However, advances in experimental techniques and computational fluid dynamics (CFD) have enabled researchers to quantify these hydrodynamic effects, revealing the intricate interactions between individual fish and the surrounding flow. These developments have been crucial in identifying the potential for energy savings and performance enhancement through schooling interactions, especially in carangiform swimmers \citep{liao2007review,pavlov2000patterns,li2019energetics,seo2022improved,timm2024multi}.

Despite these advances, obtaining precise measurements of the flow fields generated by individual fish within a school remains challenging. Fish within schools move unpredictably, making it difficult to control their positions or gather detailed fluid dynamic data at the individual swimmer level. Consequently, many experimental studies \citep{herskin1998energy,cooke2004activity,taguchi2011rainbow,zhang2024collective} focus on quantifying the metabolic costs of groups of fish, successfully capturing the phenomenon of energy savings but falling short of uncovering the precise mechanisms that drive it, thus limiting translational applications. Computational flow modeling studies \citep{pan2024unraveling,seo2022improved,verma2018efficient} have achieved three-dimensional, high-fidelity reconstructions of fish schooling flow fields, yet these models are constrained by the substantial computational costs associated with simulating large, complex fish schools over time. High-fidelity simulations require extensive computational resources, often relying on supercomputers. Additionally, in CFD simulations, fish position is typically prescribed and as the number of swimmers increases, the possible configurations for fish in a school expand rapidly. As a result, CFD studies are often limited to small groups or simplified models, which cannot fully capture the complex dynamics of real-world schooling behavior.
\begin{figure}
    \centering
    \includegraphics[width=\textwidth]{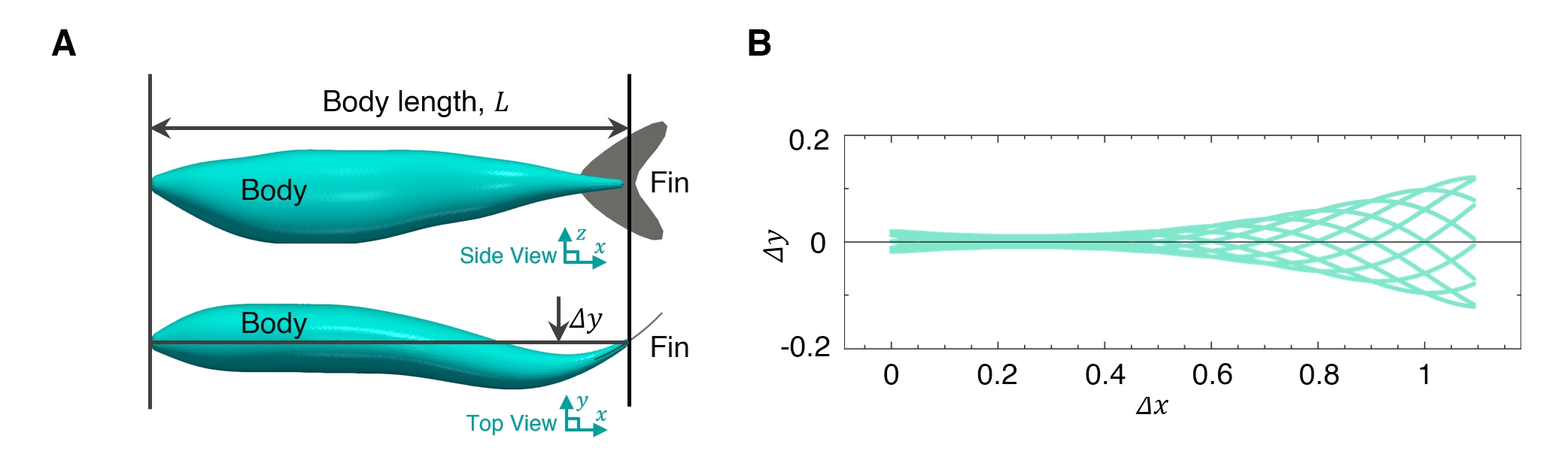}
    \caption{\textbf{3D model and centerline kinematics of a solitary fish.} \textbf{(A)} Side and top views of the simulated fish, showing the body and caudal fin. \textbf{(B)} Lateral displacement of the fish centerline over one tailbeat cycle, $\Delta t = T/10$, illustrating the kinematic motion along the axial length, $\Delta x$.}
    \label{fig:simulationConfiguration}
\end{figure}

To address these experimental and computational challenges, in this study, we propose a leading-edge vortex-based model (LEVBM) that leverages insights from solitary fish studies and flapping foil models to predict the hydrodynamic performance of trailing fish in a school. Our prior studies \citep{seo2022improved,zhou2024effect} have demonstrated the critical role of leading-edge vortices (LEVs) in the thrust generation of individual fish, particularly in the context of caudal fin dynamics (details in the next section). In parallel studies, we have validated the LEVBM for thrust generation from flapping foils \citep{raut2024hydrodynamic} and shown that the LEVBM can be used to guide the relative placement of foils in multifoil propulsors so as to maximize the gains in thrust generated from the hydrodynamic interactions between the foils \citep{Raut2025}.

In the current study, we extend this idea to ``carangiform'' body-caudal fin (BCF) swimmers, fish or fish-like swimmers that use an undulatory motion of their body and caudal fin. The LEVBM uses pre-simulated wake flow fields from a leading fish and the tailbeat kinematics of the trailing fish to evaluate the potential thrust generation of the trailing fish placed in the wake of the leading fish, without employing additional high-fidelity simulations. This enables us explore a wide parameter space, including relative position, tailbeat phase, amplitude, Reynolds number and number of fish, and gain new insights into the hydrodynamic implications for coordinated swimming in these BCF swimmers.

\section{\label{sec:methods} Methods}
\subsection{LEV-Based Model (LEVBM) for Thrust Generation}
In our previous study \citep{raut2024hydrodynamic}, we have proposed a model to predict thrust generation by a pitching and heaving foil based on its kinematics. Given that the caudal fin of BCF swimmers functions similarly to a pitching and heaving foil, this model can also be applied to predict the thrust generation by the caudal fin.

The swimming kinematics of the current BCF swimmer are given by the following equation:
\begin{equation}
\Delta y (x) = A(x
) \sin(kx - 2 \pi ft ); \quad
A(x)/L = a_0 +a_1(x/L)+a_2(x/L)^2,
\label{eq:prescrivedMotion}
\end{equation}
where $\Delta y(x)$ is the lateral displacement, $L$ is the body length, $x$ is the axial distance from the nose, $f$ is the tailbeat frequency and $A(x)$ is the amplitude modulation function. Based on literature \citep{videler1984fast} and our previous studies (\citep{seo2022improved,zhou2024effect}), the parameters are set as $k=2\pi/L$, $a_0=0.02$, $a_1=-0.08$, and $a_2=0.16$. Based on these kinematics, the heaving ($h(t)$) and pitching motion ($\theta(t)$) of the caudal fin, which is located at $x_c/L=1$ is given by:
\begin{equation}
    \begin{aligned}
    h(t) &= A(x_c) \sin(kx_c - 2 \pi ft );  \\
    \dot{h}(t) &= -2\pi f A(x_c)\cos{\left(kx_c-2\pi ft \right)}; \\
    \theta(t) &= \tan^{-1}\left[ {\partial h/\partial x}\right]_{x=x_c} \approx \tan^{-1} \left[  (kA(x_c)\cos(kx_c-2\pi ft))\right]; \\
    &= \tan^{-1} \left[ -(k/({2 \pi f}))  \dot{h}(t) \right] = -\tan^{-1} \left( \frac{\dot{h}(t)}{V_b} \right);  
    \end{aligned}
\label{h_dot}
\end{equation}
where $V_b= 2 \pi f/k$ is the wave velocity of the body undulation. The above expression for $\theta$ assumes that $dA/dx$ is small. In the LEV-based model, the Kutta-Joukowski theorem is applied to express the force normal to the  caudal fin as:
\begin{equation}
    F_N = \rho V \Gamma,
    \label{FnEquation}
\end{equation}
where $\Gamma$ represents the net circulation generated by the caudal fin, and $V$ is the net relative velocity of the fin to the flow, defined as $V = \sqrt{U^2 + \dot{h}^2}$, where $U$ is the swimming speed in the surge direction. According to our findings from the force partitioning method analysis \citep{seo2022improved,raut2024hydrodynamic}, the circulation $\Gamma$ for these flapping foils/fins is primarily due to the LEV, whose strength is proportional to the velocity component of $V$ perpendicular to the chord of the fin. The magnitude of this velocity component is related to the instantaneous effective angle of attack ($\alpha_\textrm{eff}$) on the caudal fin. Thus, we assume a proportional relationship between $\Gamma$ and $\alpha_\textrm{eff}$:
\begin{equation}
    \Gamma \propto V \sin (\alpha_\textrm{eff}).
    \label{FnAlphaEff}
\end{equation}
The instantaneous effective angle of attack, $\alpha_\textrm{eff}$, is calculated as:
\begin{equation}
    \alpha_\textrm{eff}(t) = -\tan^{-1}\left(\frac{\dot{h}(t)}{U}\right) - \theta(t)= -\tan^{-1}\left(\frac{\dot{h}(t)}{U}\right) + \tan^{-1} \left( \frac{\dot{h}(t)}{V_b} \right).
    \label{alphaEffDef}
\end{equation}

Given that the LEV-induced force is primarily determined by the relative flow velocity at the leading edge, $V$, we can express the force coefficient, $C_N$, as:
\begin{equation}
    C_N = \frac{F_N}{\frac{1}{2} \rho V_\textrm{max}^2 c},
    \label{modified_thrust_coeff}
\end{equation}
 where $c$ is the length of the caudal fin and $V_\textrm{max}$ is the maximum value of $V$ during the tail-beat cycle. Using Eq.~\ref{FnAlphaEff}, we find that $C_N \propto \sin(\alpha_\textrm{eff})$. Consequently, the thrust coefficient, $C_T$, can be expressed as:
\begin{equation}
    C_T \propto \sin(\alpha_\textrm{eff}) \sin(\theta).
    \label{eq_ctprop}
\end{equation}
We define the LEV thrust factor, $\Lambda_T$ as the mean value of the RHS of the above expression as follows:
\begin{equation}
\Lambda_T = \langle \sin(\alpha_\textrm{eff}) \sin(\theta) \rangle,
\label{eq_lambdaT}
\end{equation}
where $\langle \cdot \rangle$ represents the mean, and based on the above expression, the mean thrust coefficient $C_T$ is expected to be linearly proportional to $\Lambda_T$. Thus, using this model, the thrust can be related to the kinematics of the foil/fin via $\Lambda_T$. This linear relationship has been extensively verified for a pitching and heaving foil in a previous study \citep{raut2024hydrodynamic} where we conducted 462 distinct simulations of flapping foils with different Strouhal numbers, pitch amplitudes and locations of the pitch axis. The linear correlation over this entire range was found to match with a $R^2$ value of 0.91, which affirmed the predictive power of the model. We employ this same model in the current study but provide additional validation of the model for the caudal fins of the BCF carangiform swimmers in a later section.

\subsection{Modeling Thrust Enhancement due to Hydrodynamic Interactions in a Fish School\label{sec:alphaT_method}}
\begin{figure}
\centering
\includegraphics[width=0.8\textwidth]{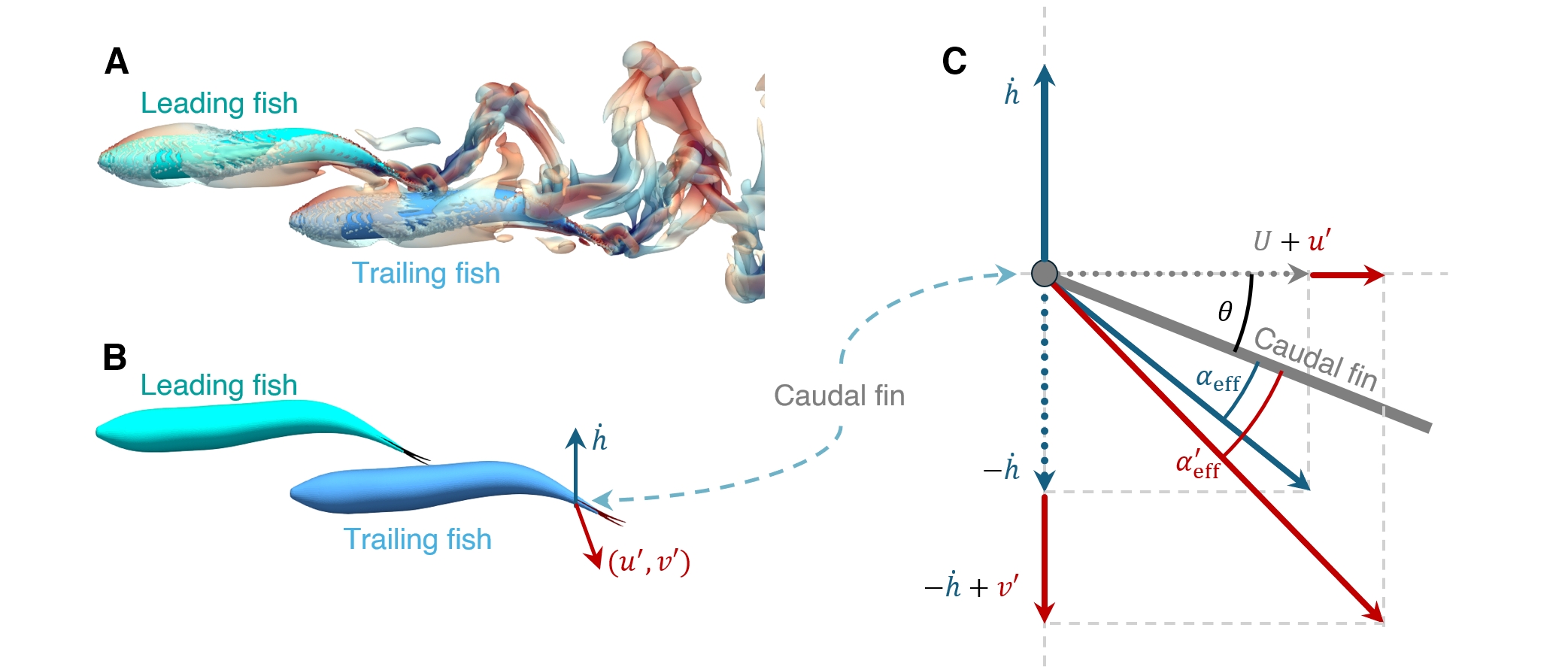}
\caption{
\textbf{Illustration of the LEV-based model used in this study.} 
\textbf{(A)} Vortex structure of a minimal school of two fish swimming in tandem, with the trailing fish positioned to interact with the wake produced by the leading fish. 
\textbf{(B)} Schematic representation showing the relative positioning of the leading and trailing fish, highlighting the motion of the caudal fin ($\dot{h}$) and flow perturbations $(u', v')$.
\textbf{(C)} Diagram of the caudal fin of the trailing fish, illustrating the effective angle of attack, $\alpha_{\text{eff}}$, and the modified angle $\alpha'_{\text{eff}}$ due to flow perturbations. The diagram highlights the influence of relative velocity components $(U + u')$ and $(-\dot{h} + v')$ on thrust generation.
}
\label{fig:alphaEff}
\end{figure}
The primary hydrodynamic distinction between
swimming alone and within a school for a fish is the ability to exploit the wake produced by other swimmers to improve its swimming performance. One significant characteristic of this wake field is the perturbation in velocity, with the lateral component being particularly dominant \citep{seo2022improved}. This lateral velocity perturbation alters the relative velocity normal to the caudal fin of a trailing fish, as depicted in Fig. \ref{fig:alphaEff}. When the local velocity perturbation $(u'v')$ are included, the effective angle-of-attack between the incident flow and the caudal fin changes from $\alpha_\textrm{eff}$ to $\alpha^\prime_\textrm{eff}$:
\begin{equation}
    \alpha_\textrm{eff}^{\prime}(t) = \tan^{-1}\left(\frac{-\dot{h}(t)+v^{\prime}(t)}{U+u^{\prime}(t)}\right) - \theta(t); \quad 
    \Lambda^{\prime}_T = \langle \sin(\alpha^{\prime}_\textrm{eff}) \sin(\theta) \rangle
    \label{alphaEffDef_prime}.
\end{equation}
This change affects the strength of the LEV generated on the caudal fin of the trailing fish, and if the movement of the caudal fin is timed appropriately with respect to this velocity perturbation, it can augment the thrust force for the trailing fish. Per this LEVBM, any improvement in thrust is proportional to:
\begin{equation}
\Delta \Lambda_T = \Lambda^{\prime}_T - \Lambda_T,
\label{eq:LambdaTPrime_LambdaT}
\end{equation}
and we use this parameter (specifically, the relative increase relative to the baseline value, i.e. $(\Delta \Lambda_T/\Lambda_T) \times 100 \%$) for quantifying the effect of hydrodynamic interactions on the thrust of trailing fish. Note that based on this expression, the change in thrust for a trailing fish whose caudal fin is located at $\left( X_T,Y_T \right)$ relative to the caudal fin of the leading fish, is a function of the wake perturbation to the velocity at that location due to the leading fish, i.e., $(u^\prime, v^\prime) = \left( u_L(X_T,t)-U, v_L(X_T,t) \right)$, where $\left( u_L(X_T,t), v_L(X_T,t) \right)$ represents the velocity field in the wake of leading fish. The fin kinematics of the trailing fish can be expressed via $\left( h_T(t), \theta_T(t), \phi_T \right) $. Thus, the thrust change for a trailing fish at any location for a given wake of a leading fish can be expressed in a functional form as: 
\begin{equation}
\Delta \Lambda_T= \Delta \Lambda_T \left[  u_L(X_T,t), v_L(X_T,t), h_T(t), \theta_T(t), \phi_T    \right]. 
\label{eq:DeltaLambdaT}
\end{equation}
It should be noted that since the velocity perturbation in the wake of the leading fish is a function of the kinematic parameters of the leading fish, the above functional relationship can also be expressed as
\begin{equation}
\Delta \Lambda_T= \Delta \Lambda_T \left[  h_L(t), \theta_L(t), \phi_L;  X_T, Y_T, h_T(t), \theta_T(t), \phi_T   \right], 
\label{eq:DeltaLambdaT_tBased}
\end{equation}
where the first three parameters depend on the leading fish and the last five parameters on the trailing fish. ``Thrust enhancement maps'' of this quantity $\Delta \Lambda_T$ will be used to interpret the results of this study. 

There are several assumptions regarding the hydrodynamics in the above model for thrust prediction of the trailing fish. Among these is the assumption that the body of the trailing fish does not affect the velocity perturbation experienced by the caudal fin. The relatively large body combined with its upstream placement relative to the caudal fin makes this an important assumption. Other notable assumptions are that the movement of the caudal fin of the trailing fish does not affect the $\left( u^\prime , v^\prime \right)$ experienced by the fin. Finally, the LEVBM  also does not account for the specific 3D shape of the caudal fin. We will examine these assumptions later in the paper.

\section{Results}
\subsection{Direct Numerical Simulation of a BCF Carangiform Swimmer}
\label{sec:results_solitary}
To resolve the flow field around the swimming fish, we use a sharp-interface, immersed boundary solver, ViCar3D \citep{mittal2008versatile}, to solve the incompressible Navier-Stokes equations with second-order finite difference in time and space. This solver has been validated extensively in previous studies of bio-locomotion flows \citep{zhou2023effect,seo2022improved,zhou2024effect,mittal2024freeman,kumar2025batwings}.

\begin{figure}
\centering
\includegraphics[width=\textwidth]{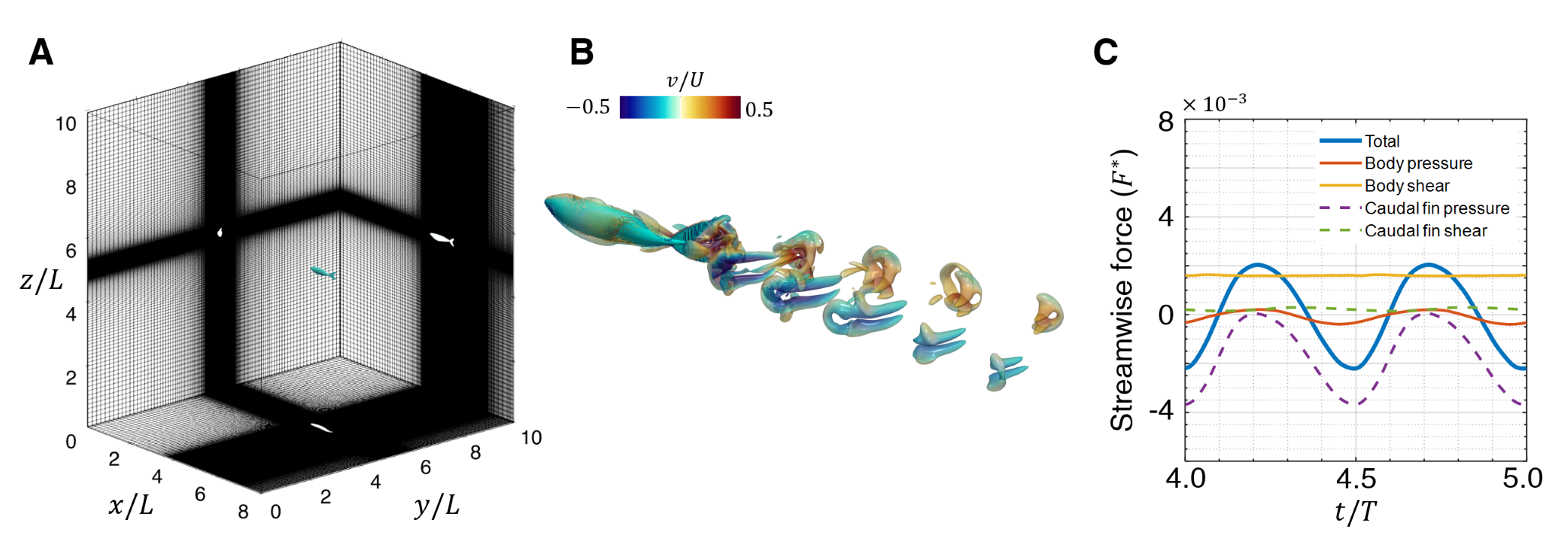}
\caption{
\textbf{Simulation of a solitary BCF swimmer.} 
\textbf{(A)} Computational domain for solitary fish swimming, with spatial dimensions shown in terms of normalized body length, $L$. The flow direction is from $-x$ to $+x$. 
\textbf{(B)} Instantaneous 3D vortex structure of the fish, visualized by the iso-surface of $Q/f^2 = 1$, colored by the lateral velocity, $v/U$.
\textbf{(C)} Pressure and viscous shear forces on the fish body and caudal fin in the streamwise direction. Forces are normalized as $F^*=F/(\rho L^4f^2)$, and $T=1/f$ is the period of tailbeat.}
\label{fig:solitaryFish}
\end{figure}
The solitary fish is tethered to a fixed location in an incoming flow in these simulations. Through trial-and-error, the incoming flow velocity is set to a value of $0.48fL$ for which, the total mean surge force on the fish is nearly zero. This models the condition where the fish is swimming at its terminal velocity. The same inflow velocity is then imposed on all the fish when the various fish school configurations are simulated. The final swimming condition corresponds to a length-based Reynolds number of $\textbf{Re}=UL/\nu=5000$. The Strouhal number based on the swimming velocity and the peak-to-peak amplitude of the caudal fin trailing-edge is $2A(x_c)f/L=0.42$. Fig.~\ref{fig:solitaryFish}(B) shows the instantaneous vortex structure of the solitary swimming fish. The tailbeat motion generates two sets of vortex loops, propagating downstream at an oblique angle to the wake centerline. The streamwise forces on the fish body are decomposed into pressure and viscous stress forces on the body and the caudal fin and plotted in Fig.~\ref{fig:solitaryFish}(C). The viscous shear drag on the body of the fish is the main source of drag, while the pressure force from the caudal fin dominates the generation of thrust, contributing to about $96\%$ of the total thrust force. 

\subsection{Thrust Enhancement Map for a Two-Fish Configuration}
\subsubsection{Generation of Thrust Enhancement Maps}
\begin{figure}
    \centering
    \includegraphics[width=1\textwidth]{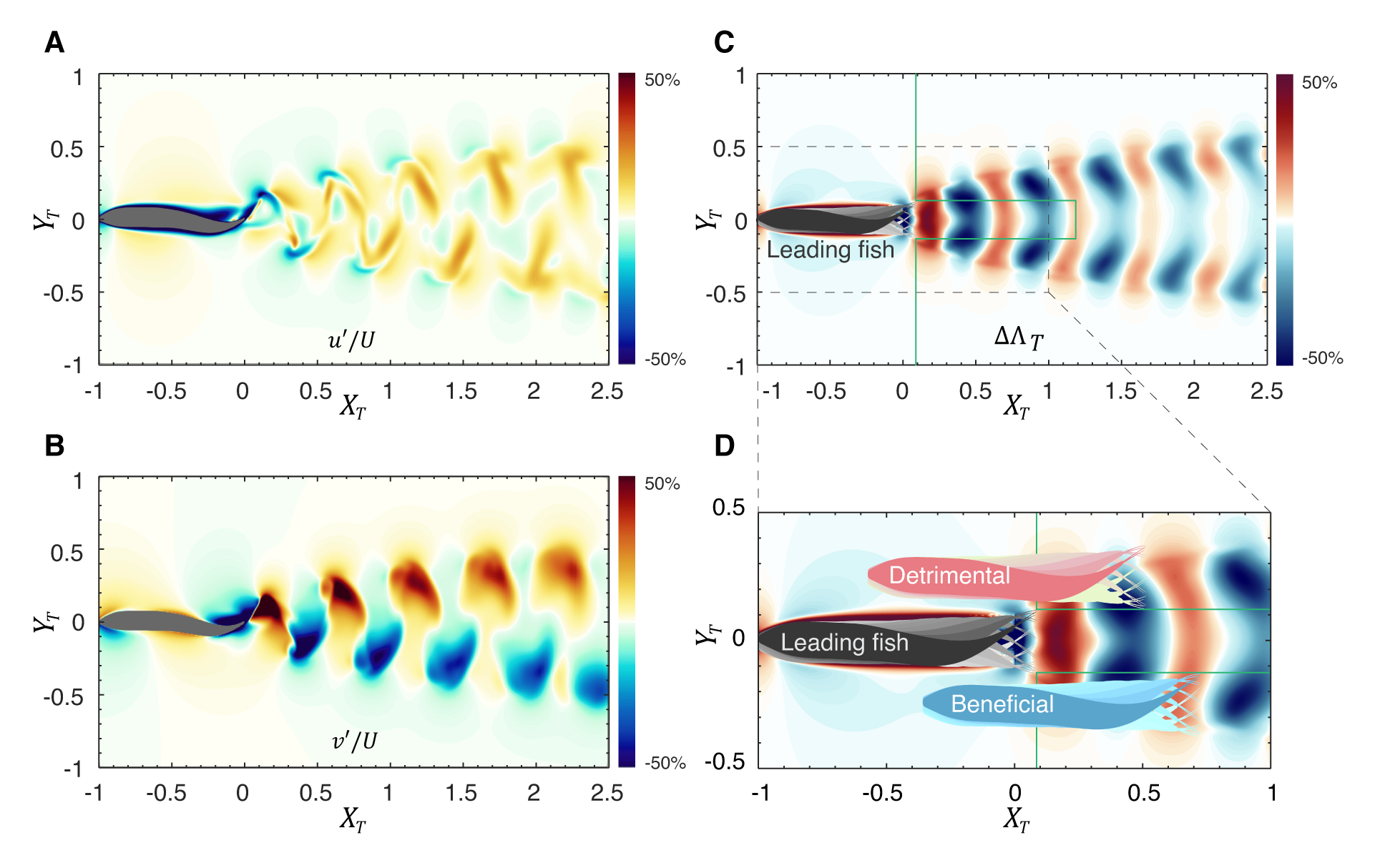}
    \caption{
        \textbf{LEVBM-based thrust enhancement map for a two-fish configuration.} 
        Instantaneous \textbf{(A)} streamwise $(u'/U)$ and \textbf{(B)} lateral $(v'/U)$ velocity components of a solitary swimmer. Velocity components are extracted from the center plane during one tailbeat cycle.  
        \textbf{(C)} $\Delta\Lambda_T$ Map ($\Delta \phi = 0$, $A = 1$, and $f = 1$) with invalid regions masked, illustrating the prediction of the thrust generation of the trailing fish.
        \textbf{(D)} Zoomed-in $\Delta \Lambda_T$ map of \textbf{(C)}, showing beneficial (red) and detrimental (blue) regions for a trailing fish. The region at the left of the green line is invalid if the trailing fish is the same size as the leading fish.
    }
    \label{fig:flowChart}
\end{figure}

We start by examining the thrust enhancement map for a two-fish configuration where both fish have exactly the same swimming kinematics in terms of frequency, amplitude, and phase. We summarize the process used to predict these hydrodynamic interactions in fish (see Fig. \ref{fig:flowChart}). Once the wake velocity field (i.e., $\left( u_L (x,y,t), v_L (x,y,t) \right)$ ) for the leading fish is obtained via a direct-numerical simulation, a trailing fish is virtually placed in this wake field at a location with its caudal fin located at $\left( X_T,Y_T \right)$ relative to the caudal fin of the leading fish on the center plane of the leading fish. The wake velocity at $\left( X_T,Y_T \right)$ is then combined with the kinematics of the caudal fin of the virtual trailing fish to estimate the interaction effect on the thrust enhancement factor $\Delta \Lambda_T$ via the expression in Eqs.~\ref{eq_ctprop}$-$\ref{eq:DeltaLambdaT_tBased}. Based on the linear relationship between $\Lambda_T$ and $C_T$ (see Eq.~\ref{eq_ctprop}), the $\Delta \Lambda_T$ is used as a surrogate for the change in thrust of the trailing fish. A thrust enhancement map is then generated by placing the virtual trailing fish at various locations within the domain with minimal computational expense.  

Fig.~\ref{fig:flowChart}(C) shows the thrust enhancement map generated based on the above procedure. We highlight a small rectangular region behind the leading fish where the trailing fish cannot be placed since this would lead to collisions between two fish. We also exclude positions of the trailing fish that would place it ahead of the leading fish. Fig.~\ref{fig:flowChart}(D) shows two notional positions of a trailing fish on the thrust enhancement map that would correspond to either a beneficial or a detrimental interaction.  In Fig.~\ref{fig:flowChart}(D), the blue trailing fish is positioned with its caudal fin within the positive $\Delta \Lambda_T$ region and would benefit from a constructive interaction with the leading fish’s wake, generating more thrust than a solitary swimmer. In contrast, the pink trailing fish, with its caudal fin located in a negative $\Delta \Lambda_T$ region, would experience a reduction in thrust.

\subsubsection{Verification of Thrust Enhancement Maps}
\begin{figure}
    \centering
        \centering 
        \includegraphics[width=1\textwidth]{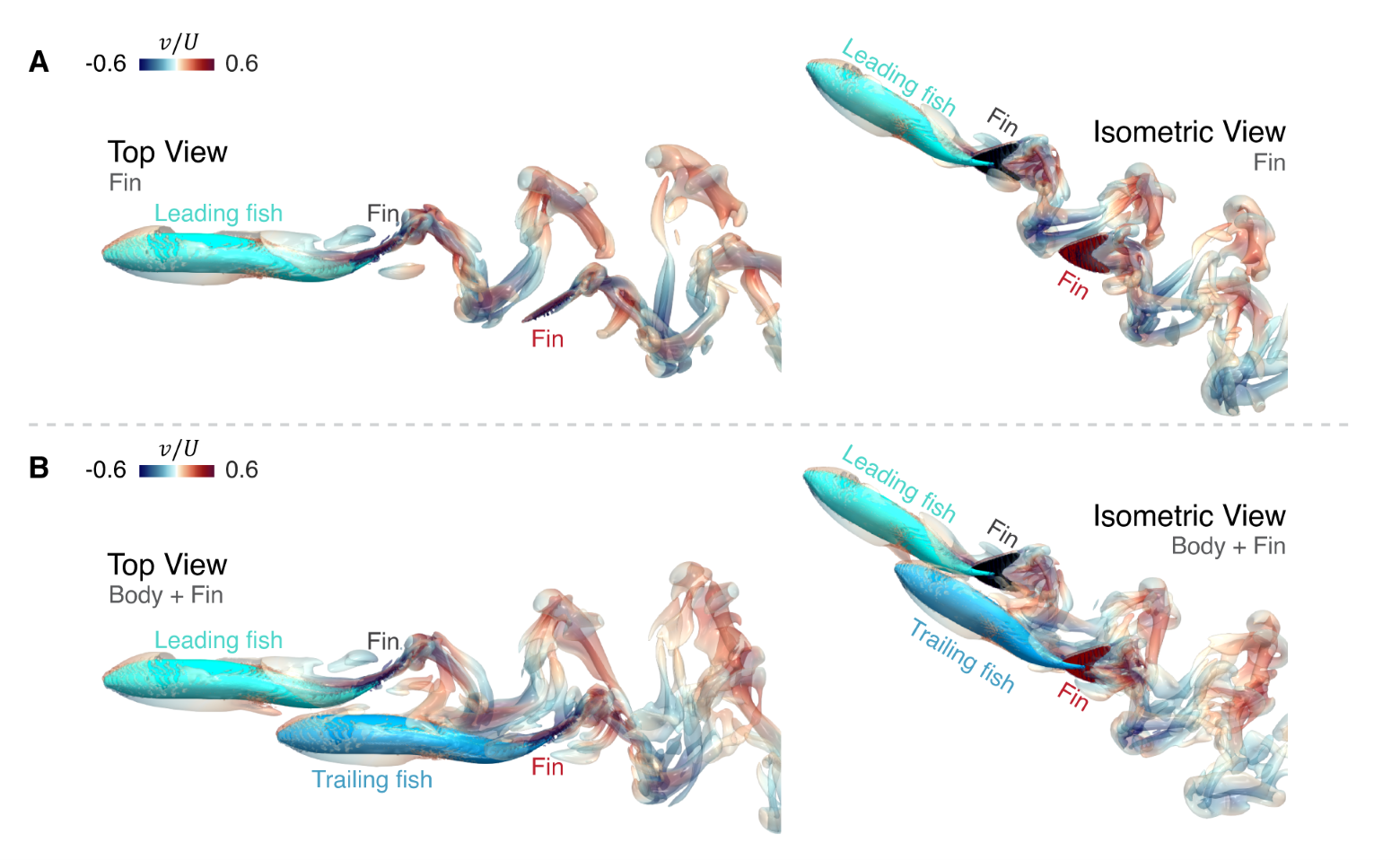}
    \caption{
        \textbf{Direct numerical simulations of two-fish schools.} 
        Top and isometric views of the instantaneous 3D vortex structures of two-fish schools. Trailing fish with (\textbf{(A)}: fin only, and with \textbf{(B)}: Body + fin). Vortex structures are visualized using iso-surfaces of $Q = 1f^2$ and colored by the normalized lateral velocity $(v/U)$, where $U$ is the steady swimming speed of the fish.
    }
    \label{fig:onlyFin}
\end{figure}
As pointed out earlier, several assumptions are inherent in the prediction of thrust for the trailing fish based on the LEVBM, and we have carried out comprehensive verifications of the model predictions to assess these assumptions. 

As noted above, a known limitation of the LEVBM is that it does not account for the effect of the body of the trailing fish on the wake perturbations encountered by the caudal fin of the trailing fish. Other key assumptions are neglecting the effect of the trailing fish's caudal fin on the encountered wake perturbation and the inability to account for the effect of the 3D shape of the caudal fin of the trailing fish on thrust enhancement. Thus, in the first set of DNS simulations, we exclude the body of the trailing fish (see Fig.~\ref{fig:onlyFin}(A) for the configuration) to test the effect of the latter two assumptions on the LEVBM predictions. 
\begin{figure}
    \centering
    \includegraphics[width=1\textwidth]{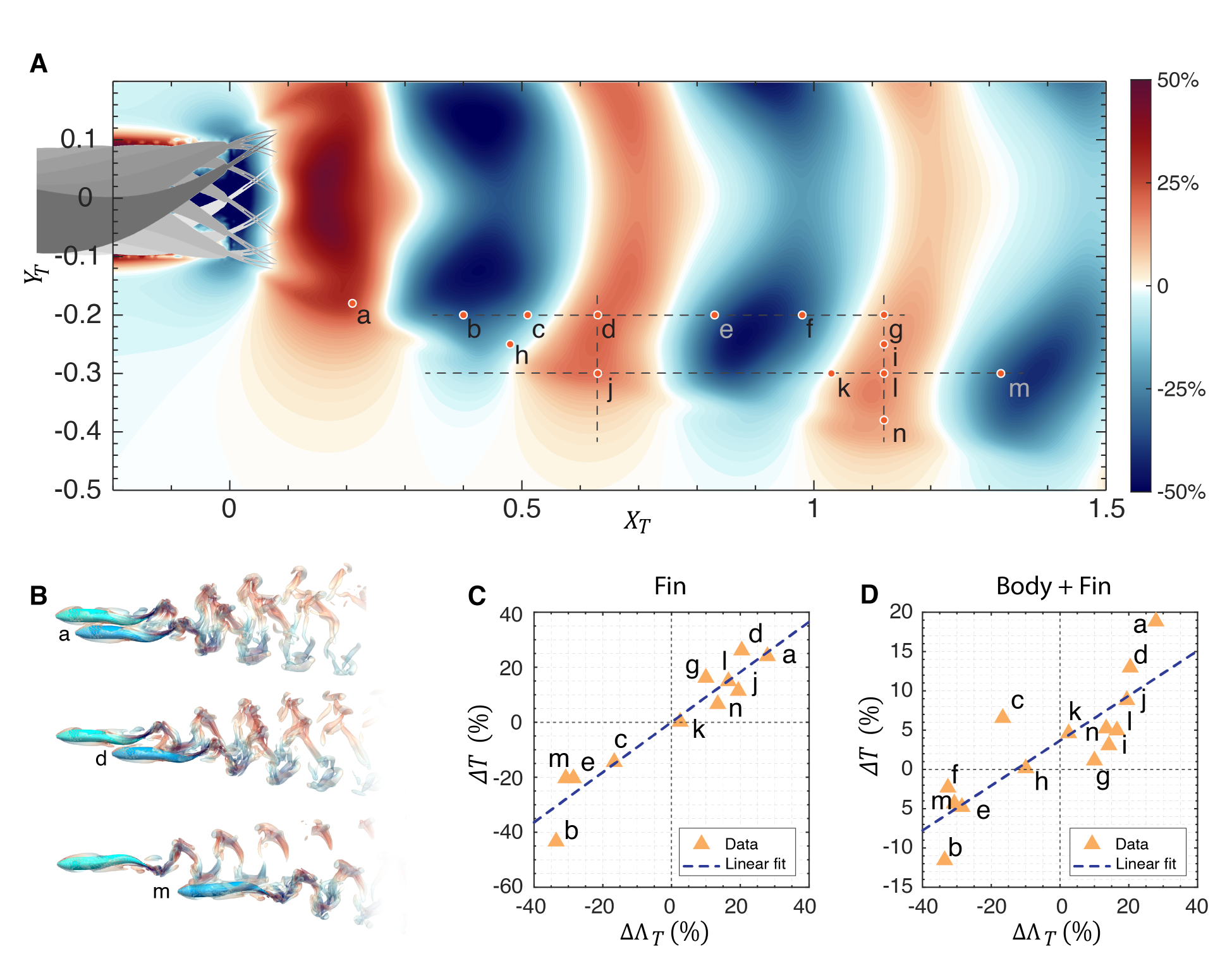}
    \caption{
        \textbf{Verification of the $\Delta \Lambda_T$ map using direct numerical simulations.} 
        \textbf{(A)} Zoomed in view of a subdomain of the $\Delta \Lambda_T$ contour map for two-fish schooling, indicating beneficial and detrimental interaction regions.  Highlighted dots show the locations of the tail of the trailing fish from positions $a$ to $n$ $(N = 14)$. 
        \textbf{(B)} Instantaneous 3D vortex structures of  two-fish schools corresponding to $a$, $d$, and $m$ in \textbf{(A)}.\textbf{(C, D)} Linear correlation between $\Delta \Lambda_T$(\%) from the LEVBM and $\Delta T$(\%) (i.e.  thrust change) from DNS, with \textbf{(C)} for fin only with an $R^2$ value of 0.9 and a corresponding best fit line of $\Delta T\%=0.91\Delta\Lambda_T\%+0.052\%$) \textbf{(D)} for body+fin with an $R^2$ value of 0.7 and a corresponding best-fit line corresponding to $\Delta T\%=0.29\Delta\Lambda_T\%+3.7\%$).}
    \label{fig:selectedPoints}
\end{figure}

For 14 selected locations of the trailing fish, which cover a range of placements with beneficial and detrimental interactions (as noted in Fig.~\ref{fig:selectedPoints}(A)), we compare the thrust enhancement predictions from the LEVBM directly to the thrust enhancement of the pressure component on the fin calculated from the DNS. Fig.~\ref{fig:selectedPoints}(C) shows the correlation between the predicted $\Delta \Lambda_T$ values and the thrust change ($\Delta T$) obtained from the DNS results corresponding to the case where the body of the trailing fish is excluded. A best-fit line between the two data sets suggests a linear relationship with a high degree of correlation ($R^2 = 0.9$). The equation of the best-fit line has a slope of 0.91 with zero intercept corresponding to 0.05\%, which signifies an excellent one-to-one relationship between the LEVBM prediction and the DNS. All in all, the above comparison suggests that the latter two assumptions discussed in the previous paragraph do not significantly deteriorate the predictions of the LEVBM, and it performs quite well despite the complex shape and effect of the caudal fin.

In the second and final stage of verification, we reintroduce the body of the trailing fish into the DNS to examine the effect of the body on the LEVBM thrust enhancement map predictions, and Fig.~\ref{fig:selectedPoints}(D) shows the correlation between the DNS estimation of the pressure thrust on the fin and the LEVBM predictions. With the body of the trailing fish included, the linear correlation reduces to a $R^2$ of 0.7, which suggests that while the presence of the fish body does diminish the predictive power of the LEVBM, the linear correlation remains acceptable, and the model is still useful for predicting the thrust changes for fish schools. We also note that the slope and intercept of the best-fit line is 0.29 and 3.7\%, respectively. This significant reduction in slope from the expected value of unity indicates that the body acts to diminish the effect of the hydrodynamic interactions on the thrust of the caudal fin. This is likely due to the fact that the most significant effect on the effective angle-of-attack of the trailing caudal fin is via the perturbation in the \emph{lateral} velocity, and the body acts as a ``wall'' and diminishes these lateral perturbations of vortex structure that convects past it to the fin. This effect of the body would be an interesting issue to explore in a future study.

\subsubsection{Observations Regarding the Topology of the Thrust Enhancement Maps}
 The periodic distribution of positive and negative regions of the $\Delta \Lambda_T$ in the streamwise direction is a result of the alternating shedding of vortices from the leading fish's tail, which convects downstream with the flow and drives the velocity perturbation pattern in the wake. As shown by \cite{seo2022improved}, the thrust enhancement is associated with the effective phase difference between the tail beats of the leading and trailing fish $\Delta \phi_\mathrm{eff}$, which is estimated by the following expression:
\begin{equation}
    \Delta \phi_\mathrm{eff} = \left( \phi_T- \phi_L \right) - \frac{2\pi X_T}{\lambda}.
\end{equation}
Here, $\phi_T$ and $\phi_L$ are the tailbeat phases of the trailing and leading fish, respectively, and $\lambda$ represents the wavelength of the velocity perturbation in the wake, which may be estimated as $\lambda \approx U/f_L$, where $U$ is the swimming speed and $f_L$ is the tailbeat frequency of the leading fish. The effective phase difference is directly related to the phase difference between $-\dot{h}_T(t)$ and $v_L^{\prime}(X_T,t)$ in Eq.~\ref{alphaEffDef_prime}, and it determines the changes in $\alpha_\text{eff}$. This suggests that the phase difference between the tail beats of the two fish, as well as the trailing fish location $X_T$ and the tail beat frequency $f_L$ of the leading fish are all factors that affect the thrust enhancement map.

\begin{figure}
    \centering
    \includegraphics[width=1\linewidth]{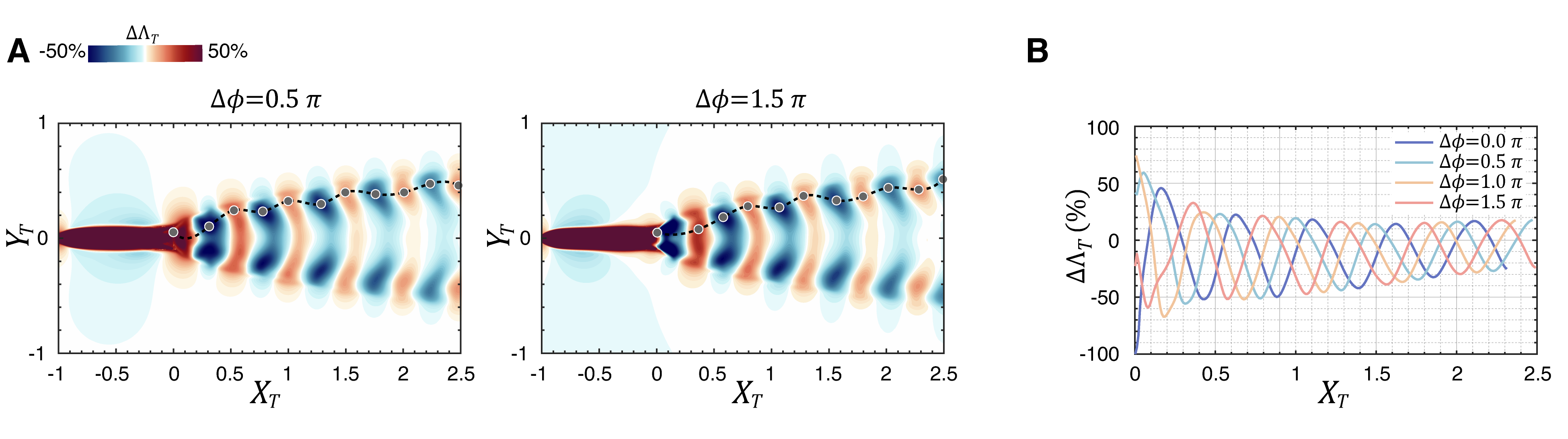}
    \caption{\textbf{$\Delta \Lambda_T$ maps for two-fish schooling at different tailbeat phases.} 
        \textbf{(A)} $\Delta \Lambda_T$ maps for two different phase differences $(\Delta \phi)$ between the leading and trailing fish. 
        \textbf{(B)} presents $\Delta \Lambda_T$ values along curves connecting peaks and valleys on contours of corresponding $\Delta \Lambda_T$ maps, demonstrated as gray dots and dashed lines in \textbf{A}. These profiles highlight variations in $\Delta \Lambda_T$ along the streamwise direction $(X_T)$, illustrating how phase influences the thrust enhancement of the trailing fish.}
    \label{fig:phiCollection}
\end{figure}

The above expression also suggests that the effects of differences in tail beat phase $\Delta \phi = \left( \phi_T- \phi_L \right) $ and separation $X_T$ between the two fish are essentially interchangeable in the near wake. In Fig.~\ref{fig:phiCollection}(A), thrust enhancement maps for two additional $\Delta \phi$ are shown alongside the $\Delta \Lambda_T$  plots along the locus of the extreme values of these quantities for four different choices of $\Delta \phi$. We note that the topology of the thrust enhancement maps for the different phases is very similar except for a shift in the streamwise direction. This is confirmed by the peak locus plot in Fig.~\ref{fig:phiCollection}(B), which shows that all the different cases of $\Delta \phi$ are essentially the same except for this shift. This suggests that a trailing swimmer can improve thrust at any given streamwise location by suitably adjusting its flapping phase. Similarly, for a given phase difference, the trailing swimmer can gain thrust benefit by moving to an appropriate location in the wake. 

Several other features of the thrust enhancement map are worth noting for their implication for schooling. First, there are multiple regions where the thrust is enhanced in the wake, but these regions are interspersed with regions where the thrust is \emph{reduced} due to the hydrodynamic interactions. The extreme values of thrust change are located along two oblique angles from the center of the wake that correspond well to the double-vortex loop
wake that is generated behind the flapping fin. The region near the wake center has relatively low values of $\Delta \Lambda_T$.

Fig.~\ref{fig:phiCollection}(B) shows the value of $\Delta \Lambda_T$ along the local peaks in this parameter and this pattern also decays relatively slowly in the wake. Indeed, while the peak in $\Delta \Lambda_T$ around $X_T \sim 0.5$ corresponds to a 25\% increase in $\Delta \Lambda_T$, the peak at a distance of 2.5 body-lengths is still 20\%. Thus, even in a minimal two-fish school, the trailing fish could achieve comparable propulsion benefits in many different locations within the wake of a leading fish. Second, even small changes in the movement of the leading fish would perturb the entire wake pattern and require the trailing fish to make larger adjustments in its location and/or flapping kinematics to recover to a beneficial state. 

\subsubsection{Detrimental Wake Interactions - Analysis and Implications}
As noted earlier, for each region in the wake where the thrust is enhanced due to the hydrodynamic interactions, there is an adjoining region where the thrust is \emph{reduced} due to these interactions. The presence of regions in the wake where the interactions are detrimental to the trailing fish is not surprising and was shown in our earlier work as well \cite{seo2022improved}. However, the LEVBM thrust enhancement map shows that the detrimental effects exceed the beneficial effects, and they also extend over larger regions of the wake than the beneficial regions. This feature is unexpected and we therefore examine this in more detail. Fig.~\ref{fig:selectedPoints}(C), which shows the correlation between the LEVBM and the DNS for the cases where the body of the trailing fish is excluded, confirms this bias towards detrimental interactions since the negative peaks in relative thrust reach -43.3\% whereas the positive peaks are limited to +26.1\%. Thus, the negative bias is not an erroneous prediction from the LEVBM but is confirmed by the DNS. We now examine the flow physics that results in this negative bias.

For a fish swimming in the wake of another fish, the flow perturbations due to the wake vortices, experienced by the caudal fin of the trailing fish will modify $\sin({\alpha_\text{eff}(t)})$ and through it, the thrust force. To understand this effect, we consider the velocity field in the wake of the leading fish in terms of a mean (denoted by ``bar") and a fluctuation (denoted by a double-prime) as follows
\begin{equation}
    \left[  u_L(x,y,t) ,  v_L(x,y,t) \right] = \left[  \bar{u}_L(x,y) ,  \bar{v}_L(x,y) \right] + \left[  u_L^{\prime\prime} (x,y,t) ,  v_L^{\prime\prime}(x,y,t) \right]
    \label{Eq:wake_vel}.
\end{equation}
We can now define an effective angle-of-attack due to the effect of the mean flow ($\bar{\alpha}_\textrm{eff}(t)$) as follows 
\begin{equation}
    \bar{\alpha}_\textrm{eff}(X_T,Y_T,t) = \tan^{-1}\left(\frac{-\dot{h}(t)+\bar{v}_L(X_T,Y_T)}{\bar{u}_L(X_T,Y_T)}\right) - \theta(t); \quad 
    \label{alphaEff_mean_wake}
\end{equation}
Similarly, the change in the thrust factor can be decomposed into 
\begin{equation}
\Delta \Lambda_T = \Delta \bar{\Lambda}_T + \Delta \Lambda_T^{\prime \prime}
\label{eq:LambdaTPrime_LambdaT_bar}
\end{equation}
where $\Delta \bar{\Lambda}_T = \langle \sin{(\bar{\alpha}_\textrm{eff})} \sin(\theta) \rangle$ is the change in thrust factor due to the mean wake and $\Delta \Lambda_T^{\prime \prime} = \Delta \Lambda_T - \Delta \bar{\Lambda}_T$ is the remaining component that is primarily associated with the velocity fluctuation in the wake. This simple decomposition now allows us to dissect the effect of the wake on the performance of the trailing fish.

\begin{figure}
    \centering
    \includegraphics[width=1\linewidth]{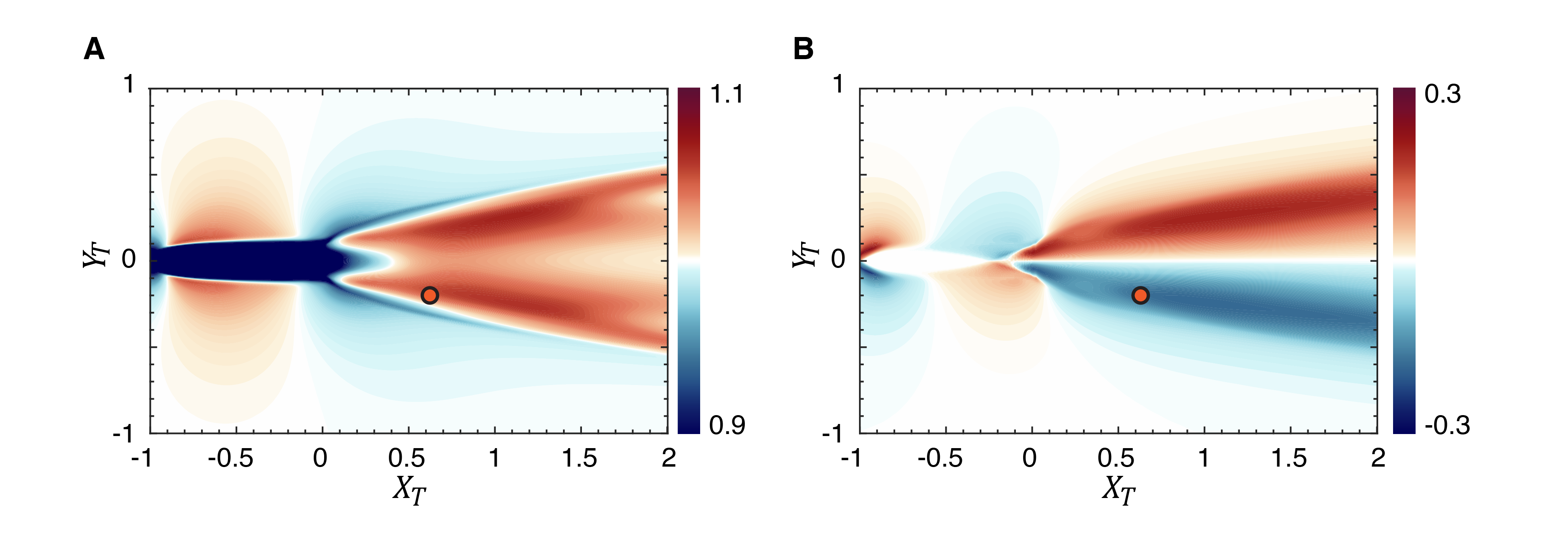}
    \caption{\textbf{Contours of the time-averaged velocity components in the wake of the solitary fish swimming .} \textbf{(A)} Contour plot of the streamwise velocity component $\bar{u}_L/U.$
    \textbf{(B)} Contour plot of the lateral velocity component $\bar{v}_L/U.$ The red dot at $(X_T, Y_T) = (0.63,-0.2)$ represents the position \emph{d} in Fig.~\ref{fig:selectedPoints}.}
\label{fig:u_bar_v_bar}
\end{figure}

Fig.~\ref{fig:u_bar_v_bar}(A) shows contour plots of $\bar{u}_L(X_T,Y_T)$ and $\bar{v}_L(X_T,Y_T)$ and we note that the mean streamwise velocity in the wake symmetric about the centerline and the values are within $\pm10 \%$ of the swimming speed.  In contrast, the mean lateral velocity is anti-symmetric about the wake centerline with values up to  $\pm 25.6 \%$ of the swimming speed. The lower part of the wake has a mean lateral component that has a negative (downwards) induced velocity, and vice-versa for the upper part of the wake. 

\begin{figure}
    \centering
    \includegraphics[width=1\linewidth]{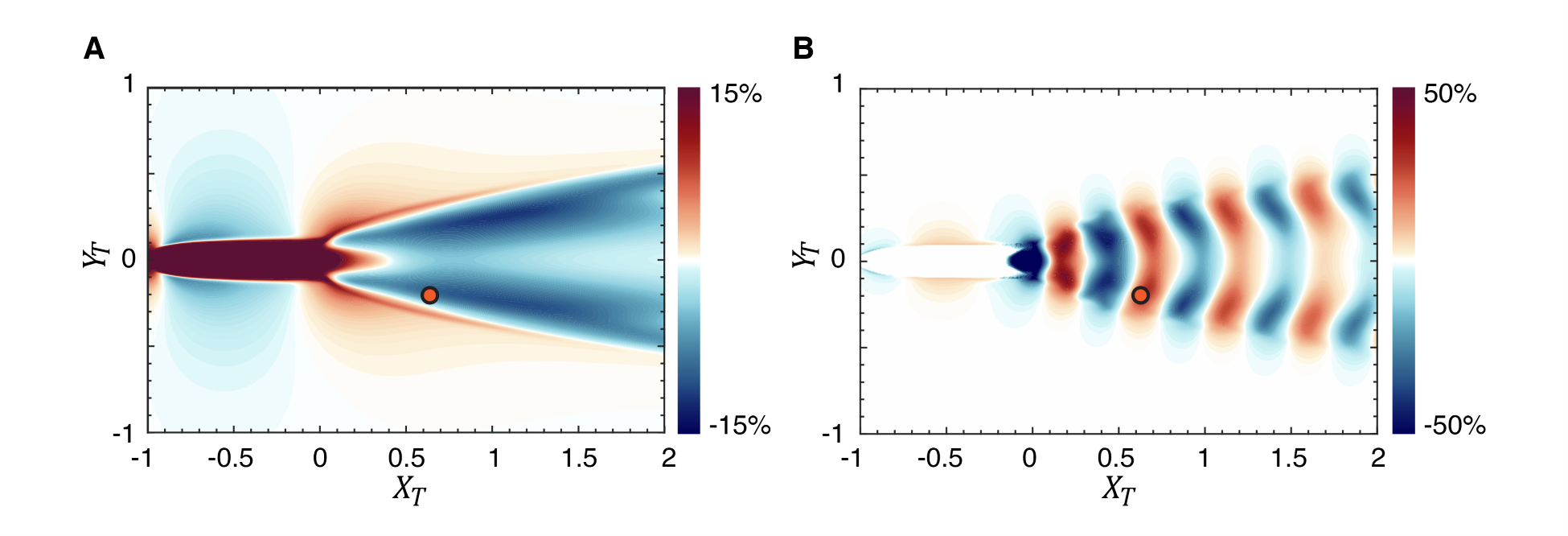}
    \caption{\textbf{(A)} $\bar{\Delta\Lambda_T}$ map computed using $\bar{u}_L$ and $\bar{v}_L$. \textbf{(B)} $\Delta \Lambda_T^{\prime \prime}$ map computed as $\Delta \Lambda_T^{\prime \prime} = \Delta \Lambda_T - \Delta \bar{\Lambda}_T$. The red dot at $(X_T, Y_T) = (0.63,-0.2)$ represents the position \emph{d} in Fig.~\ref{fig:selectedPoints}.}
\label{fig:LambdaT_meanUV}
\end{figure}

Fig.~\ref{fig:LambdaT_meanUV}(A) shows contours of $\Delta \bar{\Lambda}_T$ for the case with $\Delta  \phi=0$ and this figure shows that the mean wake has a predominantly detrimental effect on the thrust factor and would therefore result in a \emph{reduction} in the thrust of the trailing fish. In fact, the reduction in the thrust factor due to the mean flow reaches magnitudes of 10.7\%. Fig.~\ref{fig:LambdaT_meanUV}(B) shows the corresponding contour plot for $\Delta \Lambda_T^{\prime \prime}$ and it shows that with the effect of the mean flow removed, the fluctuations generate nearly similar magnitudes of beneficial and detrimental interactions. Thus, the negative bias in the thrust factor for the trailing fish is clearly due to the effect of the mean wake, which has a strong negative bias on thrust generation. 

\begin{figure}
    \centering\includegraphics[width=1\linewidth]{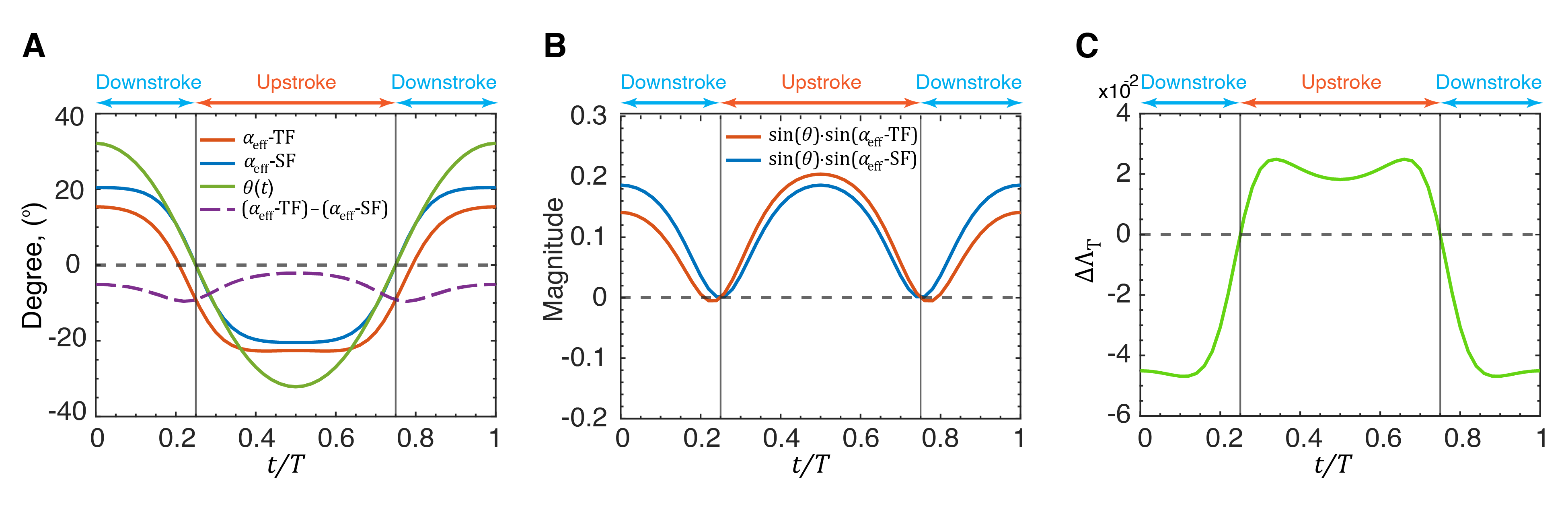}
    \caption{
    \textbf{Time variation of $\bar{\alpha}_\textrm{eff},~\theta(t), ~\bar{\Lambda}_T$, and $\Delta\bar{\Lambda}_T$ at the location \emph{d} indicted in Fig.~\ref{fig:selectedPoints}.} ``TF" and ``SF" represent ``trailing fish" and ``solitary fish," respectively. Downstroke and upstroke periods are marked on each plot.
    }
\label{fig:sin_alpha_eff_sin_theta}
\end{figure}

What remains now is to explain why the mean wake leads to a negative bias in the thrust factor. We begin by noting that $\Lambda_T$ is an inner product of $\sin({\theta(t))}$ with $\sin{(\alpha_\text{eff}(t))}$ (see Eq.~\ref{eq_lambdaT}) and large positive values of $\Lambda_T$ are generated when these two periodic functions are similar to each other in shape and phase.  We plot the $\alpha_\textrm{eff}(t)$ for a solitary fish (SF) and the $\bar{\alpha}_\textrm{eff}(t)$ for a trailing fish (TF) in two-fish configurations at location $(X_T=0.63,Y_T=-0.2)$, which is location \emph{d} on Fig.~\ref{fig:selectedPoints}(A) and which corresponds to a location close to the peak of thrust enhancement for the trailing fish. Also plotted is the pitch angle $\theta(t)$, which is the same for the two fish. As Fig.~\ref{fig:sin_alpha_eff_sin_theta}(A) shows, for the solitary fish, we find that $\alpha_\textrm{eff}(t)$ and $\theta(t)$ are very much in phase as evidenced by the fact that they cross the abscissa at the exact locations. This is not a coincidence since both functions result from the same BCF motion of the fish. In fact, as evident from Eqs.~\ref{h_dot} and \ref{alphaEffDef}, for BCF motion, $\alpha_\textrm{eff}(t)$ and $\theta(t)$ are both related to the arctangent of $\dot{h}$ and are therefore expected to be in phase.  Thus, the simple sinusoidal flapping motion of the caudal fin generates temporal profiles of $\sin({\theta(t)})$ and $\sin({\alpha_\text{eff}(t)})$ that are intrinsically well-suited for thrust generation. We note that BCF swimmers in Nature have arrived at this swimming motion through hundreds of millions of years of evolution, and it would, in fact, be puzzling if this swimming motion was not well suited for thrust generation. Indeed, the thrust factor parameter that emerges from the LEVBM provides a strong theoretical underpinning and understanding of why such a swimming mode is ubiquitous in Nature.

The plot of $\bar{\alpha}_\textrm{eff}(t)$ for a trailing fish shows some interesting differences from that of a solitary fish. The entire curve for this quantity is essentially shifted downwards by values ranging from $2^\circ$ to $10^\circ$.  We note that at this location $(\bar{u}_L,\bar{v}_L)=(1.045U, -0.168U)$ and this induces a flow angle of $-9^\circ$ which is consistent with the shift in $\bar{\alpha}_\textrm{eff}(t)$. This downwards lateral velocity at this location increases the effective angle-of-attack during the upstroke, but reduces the effective angle-of-attack during the downstroke which is akin to the fin flapping with a biased pitch angle.

    Fig.~\ref{fig:sin_alpha_eff_sin_theta}(B) shows the variation of $\sin({\theta(t)}) \cdot\sin{(\alpha_\text{eff}(t))}$ for these two cases and we note that compared to the solitary fish, the trailing fish sees a significant reduction in this quantity during the downstroke and a smaller increase during the upstroke. This asymmetry is due to the fact that the downward shift of $\bar{\alpha}_\textrm{eff}(t)$ results in a phase mismatch with $\theta(t)$, thereby diminishing the product of these two functions. Thus, the net result of the mean wake is to reduce the thrust of the trailing fish. The fluctuation in the flow velocity has nearly equal potential to either increase or decrease the thrust depending on the location. However, the negative bias effect due to the mean flow results in detrimental interactions being more significant.

\subsubsection{Influence of Trailing Fish Tail Beat Amplitude on Thrust Enhancement}
In the previous section, we examined the case where the leading and trailing fish swim with identical kinematics (i.e., the same amplitude and frequency). For these cases, the hydrodynamic interaction effects are determined almost exclusively by $\Delta \phi_\mathrm{eff}$, which depends on a combination of the difference in tail beat phases and the distance between the two fish. However, depending on its position in the wake, a swimmer in the wake of a leading swimmer will experience changes in thrust (and therefore the total surge force), which would lead to acceleration or deceleration of the swimmer. One way to maintain the position at a beneficial location or to move from a detrimental position to a beneficial position is via modification in tailbeat amplitude. Indeed, a swimmer in a beneficial location could reduce their power expenditure while maintaining position by reducing their tailbeat amplitude. It is, therefore, of interest to examine the thrust enhancement map for a fish that is swimming in the wake of another fish using a tailbeat amplitude that is different from the leading fish. The current LEVBM provides the opportunity to easily examine this question since the kinematics of the trailing fish can be modified without requiring any additional high-fidelity simulations.

\begin{figure}
    \centering
    \includegraphics[width=1\textwidth]{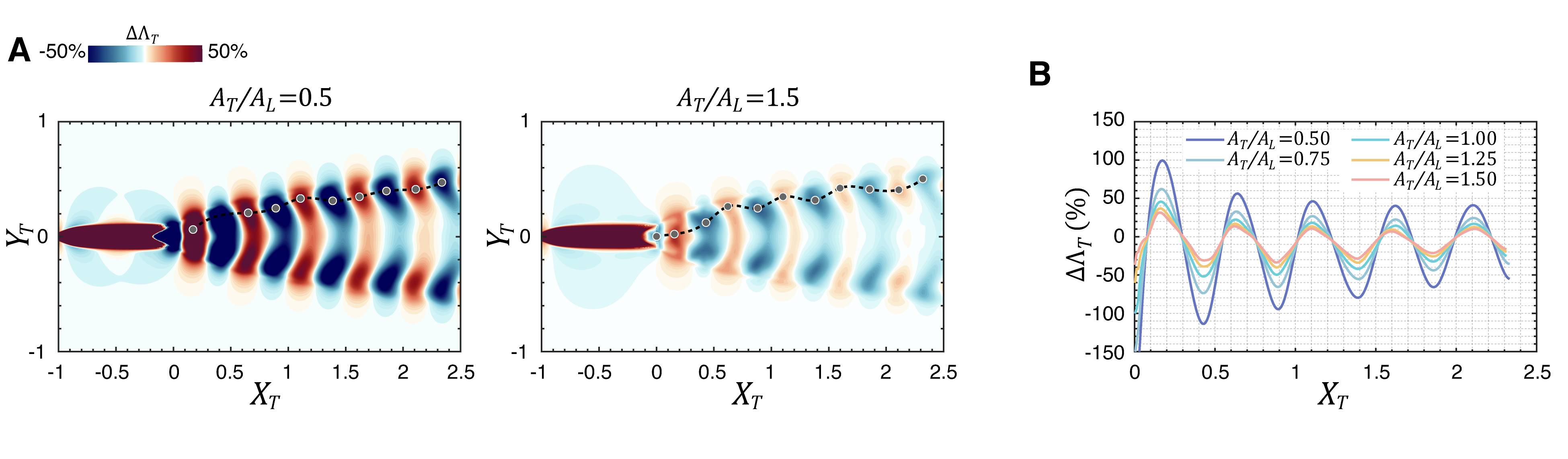}
    \caption{\textbf{$\Delta \Lambda_T$ maps for two-fish schooling with different tailbeat amplitudes.} 
        \textbf{(A)} $\Delta \Lambda_T$ maps for two different tailbeat amplitude, $A,$ of the trailing fish.
        \textbf{(B)} presents $\Delta \Lambda_T$ values along curves connecting peaks and valleys on contours of corresponding $\Delta \Lambda_T$ maps, demonstrated as gray dots and dashed lines in \textbf{A}. These profiles highlight variations in $\Delta \Lambda_T$ along the streamwise direction $(X_T)$, illustrating how amplitude influences the thrust generation.}
    \label{fig:ampChanges}
\end{figure}
Fig. \ref{fig:ampChanges} shows the effect of the tail beat amplitude of the trailing fish on thrust enhancement for the trailing fish. The $\Delta \Lambda_T$ maps for $A_T/A_L = 0.5$ and $A_T/A_L = 1.5$ reveal that the topology of the thrust enhancement map is similar to that for the same amplitude but a smaller flapping amplitude in the trailing fish ($A_T/A_L=0.5$) leads to larger relative percentage increase in the $\Delta \Lambda_T$ values. This is expected since for a trailing fish with a reduced tail beat amplitude, the wake perturbations $(u^\prime_L,v^\prime_L)$ from the leading fish’s wake are more significant relative to the trailing fish’s fin velocity, which is itself proportional to $\dot{h}_T$. Consequently, the effective angle-of-attack is modified more significantly by the wake interaction, leading to greater changes in $\Delta \Lambda_T$. Conversely, with higher tail beat amplitudes, the trailing fish’s fin motion dominates, thereby reducing the influence of the perturbation from the wake vortices. Thus, the current analysis shows that a difference in amplitude between the two fish does not fundamentally change the flow physics of thrust enhancement, and this can serve as a viable strategy for maintaining or reaching a beneficial location in the wake.

\subsubsection{Effect of Reynolds number}
A Reynolds number of 5000 corresponds to a 2-5 cm caudal fin swimmer (such as a Giant danio) swimming at O(1) BL/s (body length per second), and as we have seen that the wake at these low Reynolds number is well organized and highly periodic. It is of interest to see if the thrust enhancement maps are affected by an increase in Reynolds numbers, which would result in a complex, non-periodic wake with transitional/turbulent flow characteristics. To examine this, we employ our solver to compute the flow past the swimmer at a Reynolds number of 50,000. These simulations are carried out on an extensive 233 million point grid, which is chosen after a grid convergence study (\citep{mittal2024freeman}). To simulate a condition corresponding to steady swimming at a terminal speed, we have conducted a series of simulations at different freestream velocities and selected a velocity for which the mean thrust and drag are nearly balanced out, and the net mean hydrodynamic force on the swimmer is almost zero. This condition for Re=50,000 corresponds to a Strouhal number of 0.28, and it nominally represents a ~15-20 cm carangiform swimmer such as a medium-sized trout.

\begin{figure}
    \centering
    \includegraphics[width=\textwidth]{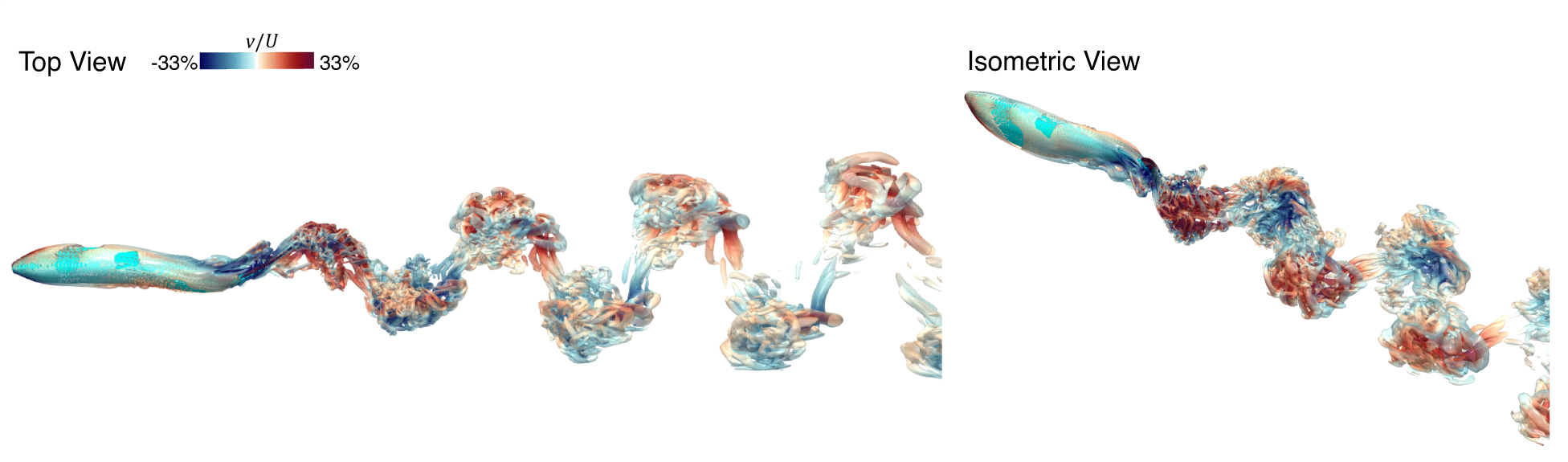}
    \caption{\textbf{Instantaneous 3D vortex structure of solitary fish swimming at Re=50,000.}
    Structures are visualized by the iso-surface of $Q/f^2 = 1$, colored by the lateral velocity, $v/U$.
            }
    \label{fig:Re50k_vorQ}
\end{figure}
Fig.~\ref{fig:Re50k_vorQ} shows two views of the vortex structures in the wake, and we note that while the wake still exhibits the characteristic oblique dual-vortex street structure, the flow is significantly more complicated with a wide range of smaller vortex structures that are a result of various instabilities in the wake.
    
\begin{figure}
    \centering
\includegraphics[width=1\textwidth]{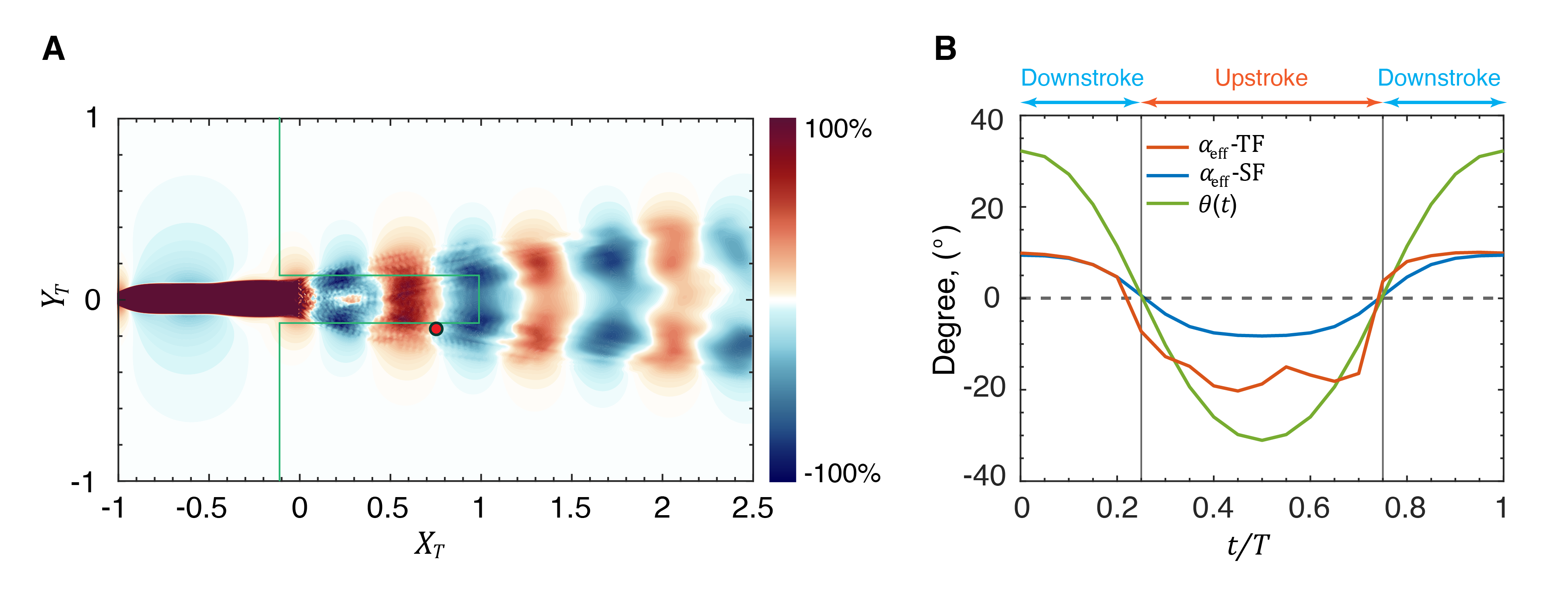}
\caption{\textbf{$\Delta\Lambda_T$ map and time variations of $\alpha_\textrm{eff}$ along with $\theta(t)$ of solitary fish swimming at Re=50,000.}
    \textbf{(A)} The $\Delta\Lambda_T$ map. The red dot at $(X_T,Y_T)=(0.95,-0.17)$ represents one location of peak values on this $\Delta\Lambda_T$ map. The region at the left of the green line is invalid if the trailing fish is the same size as the leading fish.
    \textbf{(B)} Time variations of $\alpha_\textrm{eff}$ and $\theta(t)$ at the location marked in \textbf{(A)}. ``TF" and ``SF" represent ``trailing fish" and ``solitary fish," respectively. Downstroke and upstroke periods are marked on the plot.}
    \label{fig:Re50k_lambdaT_map}
\end{figure}
This lack of strict periodicity in the wake could impact the ability of a trailing swimmer to harness the induced velocities from the wake to enhance its LEV and thrust. To examine this issue, the wake velocity field from this simulation is extracted and used to generate a thrust enhancement map of a swimmer in the wake. Fig.~\ref{fig:Re50k_lambdaT_map}(A) shows the thrust enhancement map for a swimmer that is swimming with kinematics and phases that are identical to the leading swimmer. Remarkably, we find that despite the one order-of-magnitude increase in the Reynolds number and the significantly increased complexity of the wake, the thrust enhancement map looks very similar in form to that at the lower Reynolds number. Fig.~\ref{fig:Re50k_lambdaT_map}(B) shows $\alpha_\textrm{eff}$ for a trailing fish at a beneficial location (marked in Fig.~\ref{fig:Re50k_lambdaT_map}(A)) along with the  $\alpha_\textrm{eff}$ for a solitary fish and $\theta(t)$. We observe from this figure that while $\alpha_\textrm{eff}$ for the trailing fish reflects the oscillatory and stochastic nature of the wake flow at this higher Reynolds number, there is an increase in the amplitude of this quantity, which ultimately results in a larger value of $\Delta \Lambda_T$ at this location. 
All of these observations indicate that the flow physics associated with wake-induced enhancement of the LEV is quite robust and effective for caudal fin swimmers over a large range of scales.

\subsection{Extension to Larger Schools}
\subsubsection{Schools with Three Fish}
\begin{figure}
\includegraphics[width=1\textwidth]{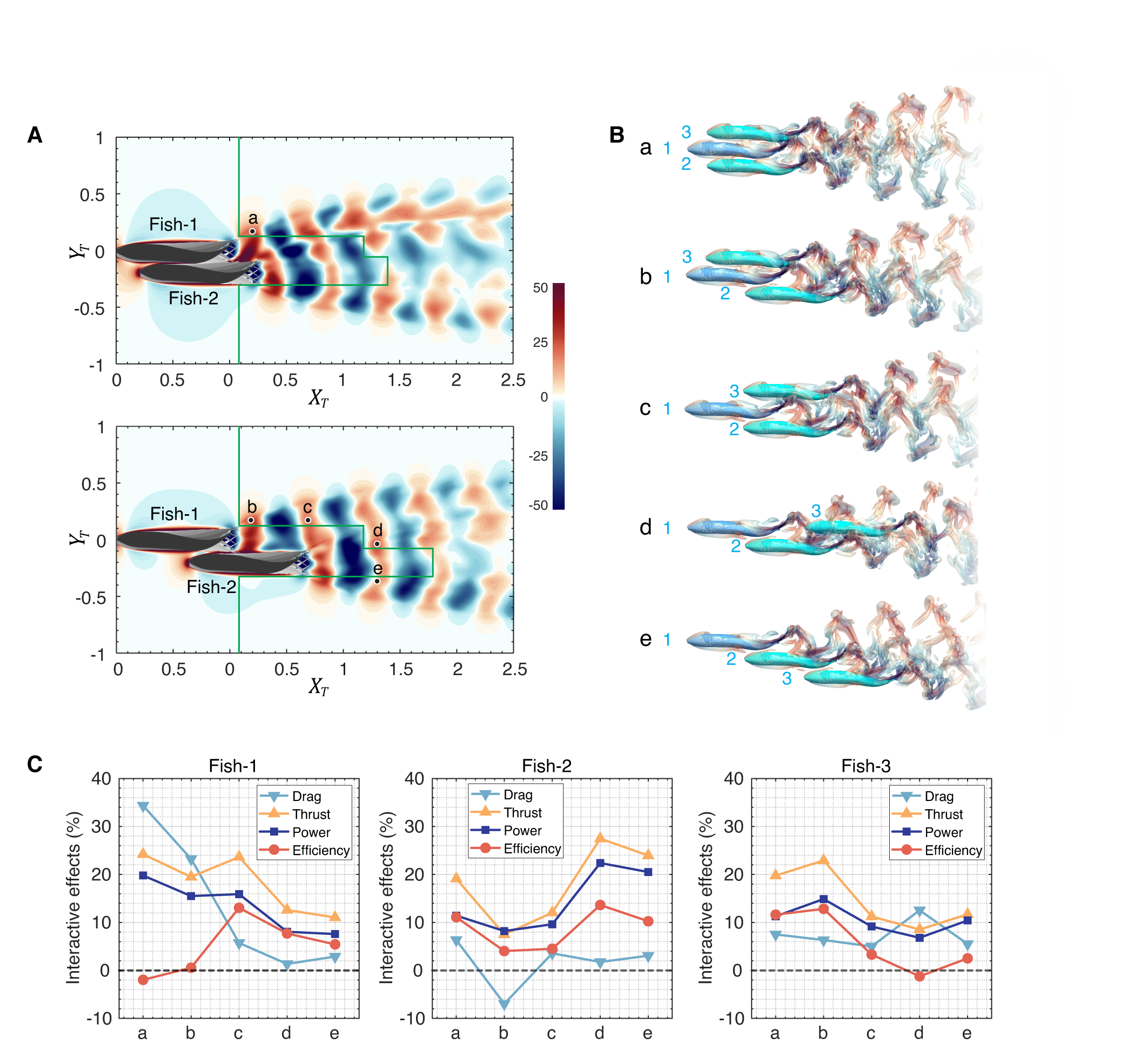}
    \caption{
        \textbf{Predicted $\Delta \Lambda_T$ map for the third fish in a three-fish school.} 
        \textbf{(A)} $\Delta \Lambda_T$ map indicating the optimal positions $(a, b, c, d, e)$ for a third fish in the wake of two leading fish with beneficial zones marked. The region to the left of the green line is invalid if the trailing fish is the same size as the leading fish.
        \textbf{(B)} Direct numerical simulations of three-fish schools. Instantaneous 3D vortex structures of three-fish schools corresponding to the optimal positions, $a-e$, in \textbf{(A)}.
        \textbf{(C)} Comparative plots of interactive effects (drag, thrust, power, and efficiency) for each fish in configurations $(a, b, c, d, e)$. Optimal positions, such as $a$ and $b$, enhance thrust and swimming efficiency for the third fish.}
    \label{fig:threeFishMaps}
\end{figure}

Based on the successful demonstration of the use of $\Delta \Lambda_T$ map for two-fish schools, we have applied the same methodology to predict beneficial configurations for three-fish schools, where the three are swimming with identical swimming strokes. For this, we first performed direct numerical simulations of two different optimal two-fish configurations, specifically $a$ and $d$ in Fig. \ref{fig:selectedPoints}(B)) and obtained their wake velocity fields. Using these velocity fields, beneficial locations for the third fish are explored by computing the respective $\Delta \Lambda_T$ maps (Fig. \ref{fig:threeFishMaps}(A)) for these cases. We note that there are many other two-fish configurations that could be examined, but we cannot explore all of them using DNS. The two configurations chosen here are however distinct enough that analysis of these two should provide general insights into the application of the LEVBM to larger fish schools.

Fig.~\ref{fig:flowChart}(C) shows the thrust enhancement maps for the two two-fish configurations, and we note that while the overall pattern of benefit and detriment is observed, the thrust enhancement maps have a significantly more complex topology. 
As shown in Fig. \ref{fig:threeFishMaps}(A), the third fish can be positioned in many beneficial locations. We select five locations: 3-Fish-a to 3-Fish-e for a detailed analysis. Fig. \ref{fig:threeFishMaps}(B) illustrates the schematic of each configuration. Placing the third fish in position $a-c$ results in the well-documented diamond formation, which has been the subject of many previous studies \citep{liao2007review,pavlov2000patterns,timm2024multi}. Alternatively, positioning the third fish at $e$ leads to a diagonal configuration, theoretically allowing for further extension into larger schooling formations.

To verify the configurations suggested by the $\Delta \Lambda_T$ maps for the third fish, we have conducted DNS of those configurations illustrated in Fig. \ref{fig:threeFishMaps}(B), and the metric related to the hydrodynamic performance of each swimmer is shown in Fig. \ref{fig:threeFishMaps}(C). The availability of the DNS for these configurations allows us to perform a more comprehensive assessment of the swimming performance of the fish, including a quantification of the effect of schooling on drag, power, and efficiency for all the fish in these configurations. We note that these additional quantities are unavailable from the LEVBM-based thrust enhancement maps.

The plots show that the thrust for every fish in all cases is enhanced, with the highest enhancement reaching 28\% ($d$, Fish-2, in Fig. \ref{fig:threeFishMaps}(C)). Noticeably, as we selected specific beneficial configurations from optimal two-fish schools, Fish-2, and even the leading fish, Fish-1, still attain a distinct thrust enhancement in configurations $a-e$ in Fig. \ref{fig:threeFishMaps}(C).
The interactive effects on drag reveal distinct trends depending on the swimmer's position within the school. For the leading fish, Fish-1, the drag increases significantly in compact configurations, such as $a$. As the school becomes less compact  ($d$ and $e$, for instance), the drag on Fish-1 returns close to the baseline, highlighting the influence of the trailing fish on upstream flow dynamics. This observation underscores the fact that the trailing fish can induce measurable hydrodynamic effects on the leading swimmer for compact configurations. In contrast, the drag changes for Fish-2 and Fish-3 are less pronounced, with variations remaining around 10\% across the various configurations. This suggests that the intermediate and trailing positions experience more stable drag conditions regardless of the compactness of the school.

The trend for power expenditure follows thrust, with increases observed for all fish in every configuration. The consistent alignment between thrust and power, combined with relatively stable drag changes, results in efficiency trends that follow similar patterns, except for Fish-1. For Fish-1, the significantly increased drag in configurations $a$ and $b$ negates the thrust gains, leading to an efficiency that is not much different from that of a solitary fish. A similar effect is also observed for Fish-3 in configuration $d$. Beyond these isolated effects, every configuration offers benefits in terms of improved thrust \emph{and} efficiency for at least two of the three swimmers. Some configurations are particularly beneficial for one of the fish (for instance, $c$ for Fish-1, $d$ for Fish-2, and $b$ for Fish-3.

In summary, these results demonstrate that the $\Delta \Lambda_T$ map not only reliably predicts thrust improvements across different configurations but also shows that these thrust improvements, which are associated with the enhancement of LEV on the caudal fin, are also often accompanied by concurrent improvements in efficiency.

\subsubsection{Application to Larger Schools}
In the previous sections, we have shown how LEVBM-based thrust enhancement maps provide an understanding of the effect of relative location and phase for fish in small schools comprising up to 3 swimmers. We have also shown that these maps apply over a wide range of scales. The final question that we address is - Can these thrust enhancement maps apply to larger schools, or is the flow in larger schools so complicated that it would not lend itself to the relatively simple phenomenology that forms the basis of the LEVBM thrust enhancement maps? 

\begin{figure}
    \centering
    \includegraphics[width=\textwidth]{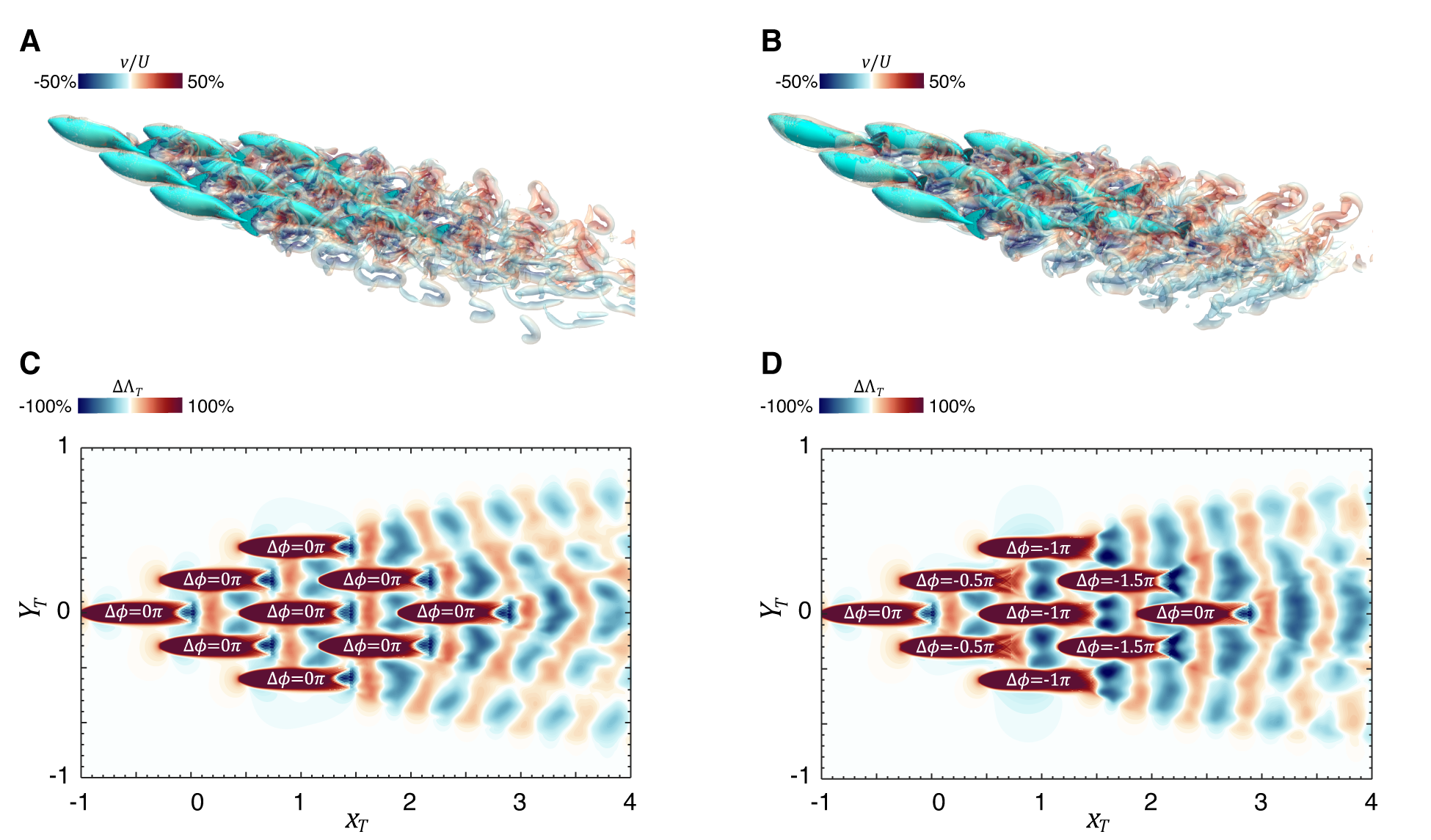}
    \caption{\textbf{Direct numerical simulations of nine-Fish school and the corresponding $\Delta\Lambda_T $ maps for a fish trailing this school.}
    \textbf{(A)} \& \textbf{(B)} Instantaneous 3D vortex structure of the nine-fish schools with synchronized and de-synchronized phases, respectively. Structures are visualized by the iso-surface of $Q/f^2 = 1$, colored by the lateral velocity, $v/U$.
    \textbf{(C)} \& \textbf{(D)}~$\Delta \Lambda_T$ maps for nine-fish schools with synchronized and de-synchronized phases, respectively. $\Delta\phi=\phi_{10}-\phi_i$, where $\phi_{10}$ is the phase of the $10^{th}$ fish.
    }
    \label{fig:9Fish_collection}
\end{figure}

To address this question, we consider two 9-fish schools that have been the subject of a previous study by us \citep{zhou2024effect,zhou2023effect}. Each individual fish in these schools corresponds exactly to the configuration examined in Sec. \ref{sec:results_solitary}. In both schools, the fish are arranged exactly in the same tight diamond configuration, but in the first school, all nine fish are swimming with the same phase, and in the second school, fish in each row are swimming with a $0.5 \pi$ phase difference from the fish in the previous row. Thus, these two schools provide two significantly distinct configurations for the application of the LEVBM. Indeed, the plots of the vortex structures for these two schools (Fig.~\ref{fig:9Fish_collection}(A) and (B))  confirm the fact that the wake flows for these two schools are not only very complex but also quite distinct. 

Thrust enhancement maps are computed for both these schools, and these are done for a trailing fish swimming with identical kinematics and a fin flapping phase that matches that of the leading fish in the schools. Fig.~\ref{fig:9Fish_collection}(C) and (D) show the thrust enhancement maps for these two schools, and we observe that while the two maps are not identical, they exhibit a surprising similarity with each other and with the thrust enhancement maps for a single leading fish. In particular, both maps are dominated by laterally oriented alternating bands of thrust enhancement and decrement. Thus, despite the tremendous complexity of the vortex wake of the 9-fish schools, the effect on the thrust of a trailing fish is quite similar to that for a single fish. The emergence of this simplicity from complexity is quite remarkable and connected with two facts: first, vortices from the various fish in the school tend to self-organize in banded structures due to mutual induction. This effect is visible in the vortex structure plots for the two schools. Second, the vorticity tends to highlight small-scale structures. However, the thrust enhancement is associated primarily with transverse velocity, which is still primarily driven by the sinusoidal movement of the caudal fin. 
\section{Summary}
The LEVBM-based thrust enhancement maps are shown to serve as a useful tool for investigating the hydrodynamics of fish schooling without incurring the high computational costs typically associated with direct numerical simulation. Unlike models based on idealized flow physics, the LEVBM-based thrust enhancement maps incorporate the essential effect of leading-edge vortex formation \citep{seo2022improved} and a detailed validation of thrust enhancement maps against direct numerical simulations provide confidence that despite its simplicity, the model provides a reasonable representation of the complex flow physics of thrust generation and enhancement associated with the LEV on the caudal fin. Leveraging the predictive capability of the model, we examined a large parameter space of locations, phase, and undulation amplitudes and frequencies of the trailing fish on its thrust enhancement. This kind of parameter sweep would not be possible using fully resolving simulations, and this capability enabled new insights into the hydrodynamics of fish schooling. 

The following are the key findings of the study:
\begin{enumerate}
\item For a trailing BCF swimmer swimming with kinematics identical to a leading BCF swimmer, there is a repeating pattern of locations in the near wake where it can derive propulsion benefits through hydrodynamic interactions with the wake of the leading fish. A change in the relative phase between the undulations of the two fish only generates a small streamwise shift in the pattern. Thus, spatial positioning and phase synchronization in fish schools are interchangeable parameters for trailing fish attempting to benefit from hydrodynamic interactions. 
\item The thrust enhancement maps generated from the LEVBM cannot account for the effects of the body of the trailing fish, but despite this, for slender-bodied swimmers such as those in the current study, the thrust enhancement maps still provide valuable predictions of the regions of thrust enhancement.
\item The thrust enhancement maps do not directly provide information regarding hydrodynamic efficiency, but the DNS show that improved thrust for the trailing fish is almost always accompanied by concurrent improvements in hydrodynamic efficiency. This is consistent with the fact that for the trailing fish, improvements in thrust are a result of LEV enhancement due to hydrodynamic interactions, and this leveraging of kinetic energy from the wake of the leading fish to amplify the growth of the LEV is power-efficient since it does not require any additional movement from the caudal fin.
\item An intriguing finding of the current study, corroborated by DNS, is that detrimental interactions (i.e., hydrodynamic interactions that reduce the thrust of the trailing fish) are not only significant, but they, in fact, exceed the magnitude of beneficial interactions. Notably, there is an alternating pattern of thrust enhancement and decrement in the wake and the dominance of the detrimental interactions is traced to the effect of the large transverse component of induced velocity in the mean wake of the leading fish. 
\item Avoidance of regions of detrimental interactions could be as crucial in the formation of schools as seeking regions of thrust enhancement. Indeed, the presence of multiple locations in the parameter space of effective phase difference where the thrust is either increased or decreased would serve to drive dynamic changes in the relative position and swimming kinematics of the trailing fish.  Taken together, all of the features of the thrust enhancement map topology described in the previous two sections provide a possible explanation for why actual fish schools, including even small-scale schools in laboratory settings, exhibit highly complex and dynamic topologies, with the fish seldom staying in, what seem to be well-organized configurations \citep{tunstrom2013collective,peterson2024Fish,zhang2024energy}.
\item A corollary to the analysis of the detrimental interaction predicted by the LEVBM-based thrust enhancement maps is that undulatory BCF swimming naturally generates a flapping motion of the caudal fin where the time-varying pitch angle and effective angle-of-attack are in phase. As per the LEVBM, this condition is very well suited (one could even consider this ``optimized'') for thrust generation. Past studies on engineered flapping foils allocated considerable effort to \emph{manually} synthesize effective angle-of-attack profiles to achieve optimal propulsion \citep{read2003forces,hover2004effect}. However BCF swimming of fish accomplishes this \emph{automatically}, in what could be considered an elegant example of embodied intelligence resulting from millions of years of evolution.
\item We find that trailing fish can swim with different undulatory amplitude than the leading fish and still reap the benefits of beneficial interactions in a manner similar to the case when the amplitudes are identical. Unsurprisingly, the relative increase in thrust decreases(increases) with increasing(decreasing) tail amplitude.
\item The thrust enhancement maps are mostly unchanged despite an order-of-magnitude increase in Reynolds number, suggesting a robustness in the LEV enhancement mechanism associated with thrust increase.
\item Although the detailed analysis in this study focused on small schools of two and three fish, the computation of thrust enhancement maps for two distinct nine-fish schools reveals a somewhat counterintuitive simplicity emerging from the complex individual wakes of the fish. This simplicity manifests as a repeating banded structure of thrust enhancement and reduction within the school's wake, which is similar to that for a single fish. These findings have intriguing implications for fish schooling, suggesting that extracting hydrodynamic benefits from the wakes of leading fish does not necessarily become more challenging as the school size increases. Similarly, this has significant implications for bio-inspired autonomous underwater vehicles (AUVs), indicating that large "schools" of such AUVs could exploit wake interactions to lower energy consumption and extend operational endurance, without requiring complex sensing and control strategies. The results from this study offer a foundation for designing and testing such systems, with the LEVBM serving as a guideline for optimizing the spacing and synchronization of AUV formations. 
\end{enumerate}

Several limitations of the study should be acknowledged. First, the simulations were conducted under idealized conditions, where the flow was assumed to be laminar, and the tethered fish maintained constant swimming speeds. In real-world scenarios, fish often swim in turbulent environments with fluctuating flow conditions. Incorporating turbulence into future models would provide a more realistic assessment of the hydrodynamic interactions in fish schools.
Additionally, the present study focused on BCF swimmers, with most of the thrust coming from the caudal fin. Future work should consider multiple fins and flexible body dynamics to better capture the full spectrum of hydrodynamic interactions, especially for other swimming modes.

Finally, the biological implications of these hydrodynamic findings remain to be fully explored. While the $\Delta \Lambda_T$ map provides trustworthy predictions for thrust generation, fish schools will likely optimize for multiple objectives simultaneously, including predator avoidance, stability, and information exchange. Investigating how these factors interact with hydrodynamics could enrich our understanding of fish schooling behavior.

\backsection[Acknowledgements and fundings]{This work is supported by ONR Grants N00014-22-1-2655 and N00014-22-1-2770 monitored by Dr. Robert Brizzolara. This work used the computational resources at the Advanced Research Computing at Hopkins (ARCH) core facility (rockfish.jhu.edu), which is supported by the AFOSR DURIP Grant FA9550-21-1-0303, and the Extreme Science and Engineering Discovery Environment (XSEDE), which is supported by National Science Foundation Grant No. ACI-1548562, through allocation number TG-CTS100002.}

\backsection[Declaration of interests]{
The authors report no conflict of interest.}

\appendix
\section{Hydrodynamic Metrics \label{sec:appendixA}}
The force and mechanical power are computed as follows through surface integrals:
\begin{equation}
\vec{F} = \int (P\vec{n} + \vec{\tau})\text{d}S, \quad
W = \int (P\vec{n} + \vec{\tau}) \cdot \vec{v} \, \text{d}S,
\label{eq:forcePower}
\end{equation}
where $\vec{n}$ is the normal unit vector pointing inside the body surface, $\vec{\tau}$ and $\vec{v}$ are the viscous stress and body velocity on the surface, respectively. Changes in force and power are defined as:
\begin{equation}
\Delta F_\mathrm{D} = F_\mathrm{D} - F_{\mathrm{D, solitary}}, \quad \Delta F_\mathrm{T} = F_\mathrm{T} - F_{\mathrm{T,solitary}}, \quad \Delta W = W - W_\text{solitary},
\label{eq:interactionSolitary}
\end{equation}
where $F_\mathrm{D}$, and $F_\mathrm{T}$ are the mean drag and thrust, calculated using the surface integral \eqref{eq:forcePower} on the body and the fin in the streamwise direction, respectively. We use the Froude efficiency \citep{zhou2018swimming,liu2017computational} ($\eta$) to quantify the hydrodynamic efficiency and the difference referring to solitary fish swimming:
\begin{equation}
\eta = \frac{\bar{F}_\mathrm{T} \tilde{U}}{\bar{W}}, \quad
\Delta \eta = \eta - \eta_\text{solitary},
\label{eq:FroudeEfficiency}
\end{equation}
where $\bar{F}_\mathrm{T}$ and $\bar{W}$ represent averaged thrust and power. $\tilde{U}$ is the adjusted swimming velocity defined as:
\begin{equation}
\tilde{U} = U + (\text{d}U/\text{d}F) \Delta F_\text{net}, \quad \Delta F_\text{net} = \Delta F_\mathrm{T} - \Delta F_\mathrm{D}.
\label{eq:adjustedUo}
\end{equation}
To correctly quantify the interactive effects, we use $\tilde{U}$ to adjust the change in drag due to the hydrodynamic interaction. The $(\text{d}U/\text{d}F)$ is obtained from the solitary fish simulations, whose Froude efficiency is 34\%, matching previous studies \citep{daghooghi2015hydrodynamic,borazjani2013fish}. 

\bibliographystyle{jfm}
\bibliography{references}
\end{document}